\documentclass[twocolumn,apj]{aastex631}
\pdfoutput=1
\usepackage{amsmath,amssymb,gensymb,color,graphicx,array,multirow,soul}
\usepackage{natbib,hyperref,multirow,placeins,textcomp}
\usepackage[outline]{contour}
\usepackage[utf8]{inputenc}
\usepackage[T1]{fontenc}
\usepackage{booktabs}
\usepackage{microtype}
\usepackage{listings}
\definecolor{olive}{rgb}{0.5, 0.5, 0.0}
\let\b\mathbf

\usepackage{expl3}

\shorttitle{Atmospheric Motion} 

\begin{document}

\title{The Atacama Cosmology Telescope: Modeling Bulk Atmospheric Motion}

\author{Thomas~W.~Morris}
\affiliation{Joseph Henry Laboratories of Physics, Jadwin Hall, Princeton University, Princeton, NJ 08544}
\author{Ricardo~Bustos}
\affiliation{Departamento de Ingeniería Eléctrica, Universidad Católica de la Santísima Concepción, Alonso de Ribera 2850, Concepción, Chile}
\author{Erminia~Calabrese}
\affiliation{School of Physics and Astronomy, Cardiff University, The Parade, 
Cardiff, Wales, UK CF24 3AA}
\author{Steve~K.~Choi}
\affiliation{Department of Physics, Cornell University, Ithaca, NY 14853}
\affiliation{Department of Astronomy, Cornell University, Ithaca, NY 14853}
\author{Adriaan~J.~Duivenvoorden}
\affiliation{Joseph Henry Laboratories of Physics, Jadwin Hall, Princeton University, Princeton, NJ 08544}
\author{Jo~Dunkley}
\affiliation{Joseph Henry Laboratories of Physics, Jadwin Hall, Princeton University, Princeton, NJ 08544}
\affiliation{Department of Astrophysical Sciences, Peyton Hall, Princeton University, Princeton, NJ 08544}
\author{Rolando~D\"{u}nner}
\affiliation{Instituto de Astrof\'isica and Centro de Astro-Ingenier\'ia, Facultad de F\`isica, Pontificia Universidad Cat\'olica de Chile, Av. Vicu\~na Mackenna 4860, 7820436 Macul, Santiago, Chile}
\author{Patricio~A.~Gallardo}
\affiliation{Kavli Institute for Cosmological Physics, University of Chicago, Chicago, IL 60637}
\author{Matthew~Hasselfield}
\affiliation{Center for Computational Astrophysics, Flatiron Institute, 162 5th Avenue, New York, NY 10010}
\author{Adam~D.~Hincks}
\affiliation{David A. Dunlap Department of Astronomy \& Astrophysics, University of Toronto, 50 St. George St., Toronto ON M5S 3H4, Canada}
\author{Tony~Mroczkowski}
\affiliation{European Southern Observatory (ESO), Karl-Schwarzschild-Strasse 2, Garching 85748, Germany}
\author{Sigurd~Naess}
\affiliation{Center for Computational Astrophysics, Flatiron Institute, 162 5th Avenue, New York, NY 10010}
\author{Michael~D. Niemack}
\affiliation{Department of Physics, Cornell University, Ithaca, NY 14853}
\affiliation{Department of Astronomy, Cornell University, Ithaca, NY 14853}
\affiliation{Kavli Institute at Cornell for Nanoscale Science, Cornell University, Ithaca, NY 14853}
\author{Lyman~A.~Page}
\affiliation{Joseph Henry Laboratories of Physics, Jadwin Hall, Princeton University, Princeton, NJ 08544}
\author{Bruce~Partridge}
\affiliation{Department of Physics and Astronomy, Haverford College,
Haverford, PA 19041}
\author{Maria~Salatino}
\affiliation{Department of Physics, Stanford University,
Stanford, CA 94025}
\affiliation{Kavli Institute for Astroparticle Physics and Cosmology,
Stanford, CA 94305}
\author{Suzanne~T.~Staggs}
\affiliation{Joseph Henry Laboratories of Physics, Jadwin Hall, Princeton University, Princeton, NJ 08544}
\author{Jesse~Treu}
\affiliation{Domain Associates, LLC}
\author{Edward~J.~Wollack}
\affiliation{NASA/Goddard Space Flight Center, Greenbelt, MD 20771}
\author{Zhilei~Xu}
\affiliation{MIT Kavli Institute, Massachusetts Institute of Technology, 77 Massachusetts Avenue, Cambridge, MA 02139}



\begin{abstract}
Fluctuating atmospheric emission is a dominant source of noise for ground-based millimeter-wave observations of the CMB temperature anisotropy at angular scales $\gtrsim 0.5^{\circ}$. We present a model of the atmosphere as a discrete set of emissive turbulent layers that move with respect to the observer with a horizontal wind velocity. 
After introducing a statistic derived from the time-lag dependent correlation function for detector pairs in an array, referred to as the pair-lag, we use this model to estimate the aggregate angular motion of the atmosphere derived from time-ordered data from the Atacama Cosmology Telescope (ACT).  We find that estimates derived from ACT's CMB observations alone agree with those derived from satellite weather data that additionally include a height-dependent horizontal wind velocity and water vapor density. We also explore the dependence of the measured atmospheric noise spectrum on the relative angle between the wind velocity and the telescope scan direction. In particular, we find that varying the scan velocity changes the noise spectrum in a predictable way. Computing the pair-lag statistic opens up new avenues for understanding how atmospheric fluctuations impact measurements of the CMB anisotropy.
\end{abstract}
\keywords{cosmic microwave background, atmospheric emission,  atmospheric modeling, turbulence}

\newpage
\section{Introduction}
\label{sec:intro}
\setcounter{footnote}{0}

The cosmic microwave background (CMB) contains a wealth of information limited only by our ability to extract it. Precisely mapping the temperature anisotropy and polarization of the CMB can help achieve numerous scientific goals, such as constraining the sum of neutrino masses, describing the distribution of dark matter, and understanding the early universe.

One of the largest challenges for ground-based telescopes is that they must observe through the atmosphere, the emission from which dominates the much fainter CMB anisotropy. In this paper we focus on the observing conditions above Cerro Toco in the Atacama Desert in northern Chile near Llano de Chajnantor, but the methods are generalizable to other sites. We demonstrate that a model based on ground-based CMB observations alone can describe the motion of the atmosphere. This paper is part of a longer term goal of quantifying how the atmosphere affects CMB anisotropy measurements so that its effects may be understood and potentially mitigated. We do not consider atmospheric polarization here, though we do note recent advances in measuring polarized scattering and emission by 
\citet{takakura/etal:2019} and \citet{petroff/etal:2020}, respectively.
\begin{figure}[thp]
\includegraphics[width=1\columnwidth]{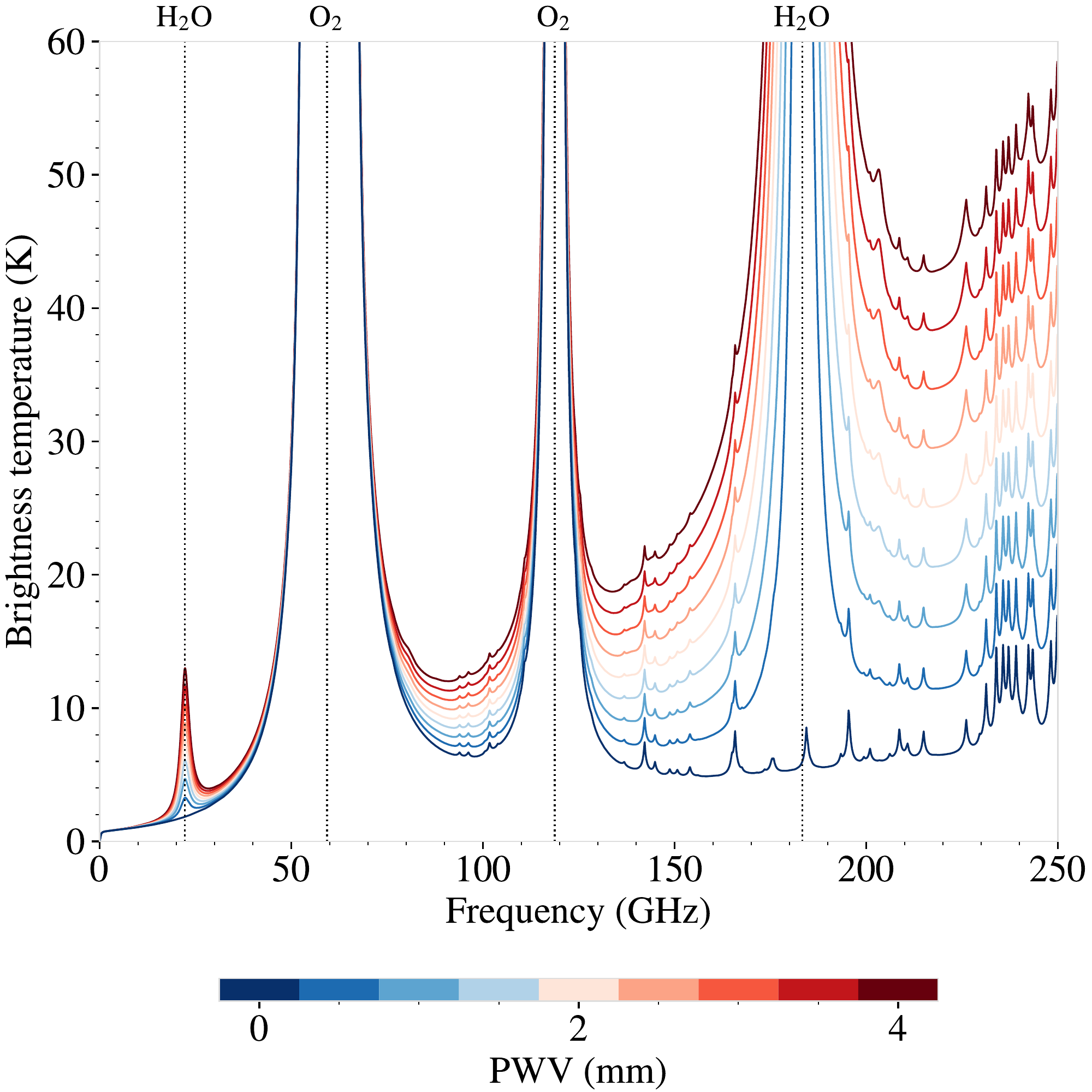}
\caption{The year-average zenith emission spectrum for the Chajnantor region for dry air and differing levels of PWV, computed using the \textit{am} software \citep{am:2018}. Fluctuations in water vapor density determine the majority of fluctuations in brightness temperature.}
\label{fig:spec}
\end{figure}

In the millimeter-wave regime, emission is dominated by two spectral lines of molecular oxygen at 60 and 120 GHz, and two water lines at 22 and 183 GHz as shown in Figure~\ref{fig:spec}; all of these ride on top of the wings of saturated water lines at higher frequencies. Of these two molecules, water is the most problematic: the concentration of water vapor\footnote{The net precipitable water vapor (PWV) emission is quantified as $\mathrm{PWV}(t)=\int_0^{\infty} \rho (h,t)dh/\rho_{H_{2}O}$ where $h$ 
is a vertical line of sight through the atmosphere, $\rho$ is the mass density of water vapor in the atmosphere, and $\rho_{H_{2}O}$ is the density of water.} is passively mixed and thus has an inhomogeneous, turbulent distribution \citep{tatarski}. This leads to variations in emission as the atmosphere moves through the line of sight. Telescopes observing the CMB through the atmosphere are thus subject to time-dependent and spatially-correlated fluctuations that both dominate the total signal, and are difficult to separate from the underlying CMB. 
\begin{figure}
\includegraphics[width=1\columnwidth]{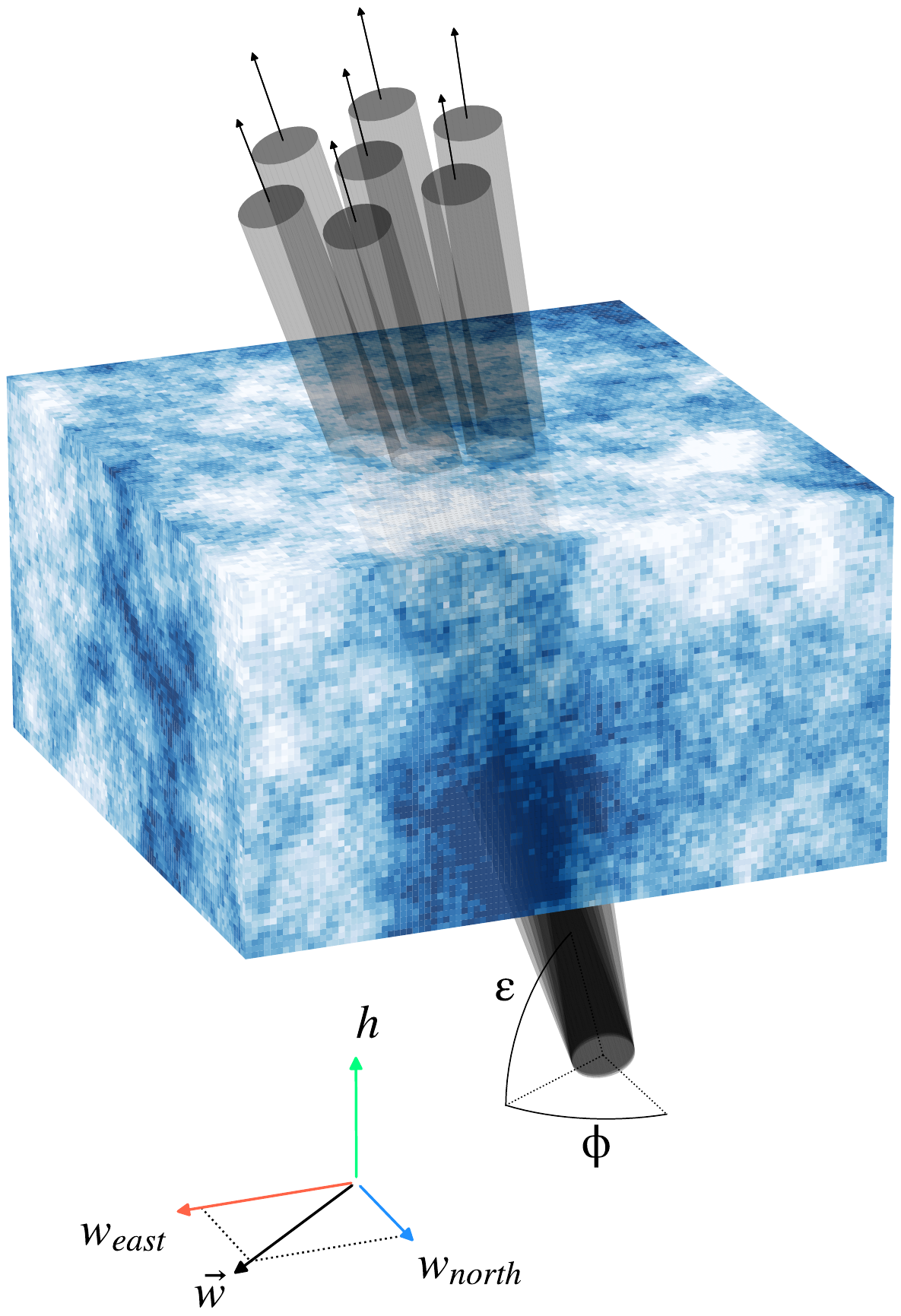}
\caption{An array of detectors peering through a section of an inhomogeneous atmosphere generated according to the covariance function derived in Section~\ref{sec:emission}. The color gradient denotes spatial variations in the time-dependent water vapor density, $\rho({\bf r},t)$. Different layers within the atmosphere can move at different velocities. The motion of the inhomogeneities with respect to the array drives the brightness fluctuations that dominate the signal of ground-based millimeter-wave telescopes. The $\vec{w}$ indicates the wind direction of, for example, the top layer. The $h$ indicates height above the observer, i.e. the ACT site.}
\label{fig:peep}
\end{figure}

Figure~\ref{fig:peep} depicts the turbulent structure of the atmosphere in three dimensions. 
Several papers such as \citet{lay:1997}, \citet{lay/halverson:2000}, \citet{sayers/etal:2010}, and
\citet{bussmann} model the atmosphere as a two-dimensional frozen sheet of turbulence moving at a constant horizontal velocity to simulate the effects of wind. This captures many aspects of the observations and is often quite effective, but cannot comprehensively describe the three-dimensional atmosphere.
Others such as \citet{church:1995} and \citet{errard/etal:2015} explicitly model the atmosphere as a continuous three-dimensional medium, which is more complete but can be computationally expensive to compare to measurements.
This paper presents a simple method for studying the motion of the three-dimensional atmosphere as it appears to a ground-based telescope by modeling it as a set of discrete two-dimensional layers. We apply the resulting model to intensity measurements from the Atacama Cosmology Telescope (ACT), and show that it recovers a useful aggregate estimate of the wind velocity that drives atmospheric fluctuations. We also show that it agrees with independent ground- and satellite-based measurements of weather parameters in the Atacama Desert. 

\section{Data Sources}
\begin{figure}[htp]
\includegraphics[width=1\columnwidth]{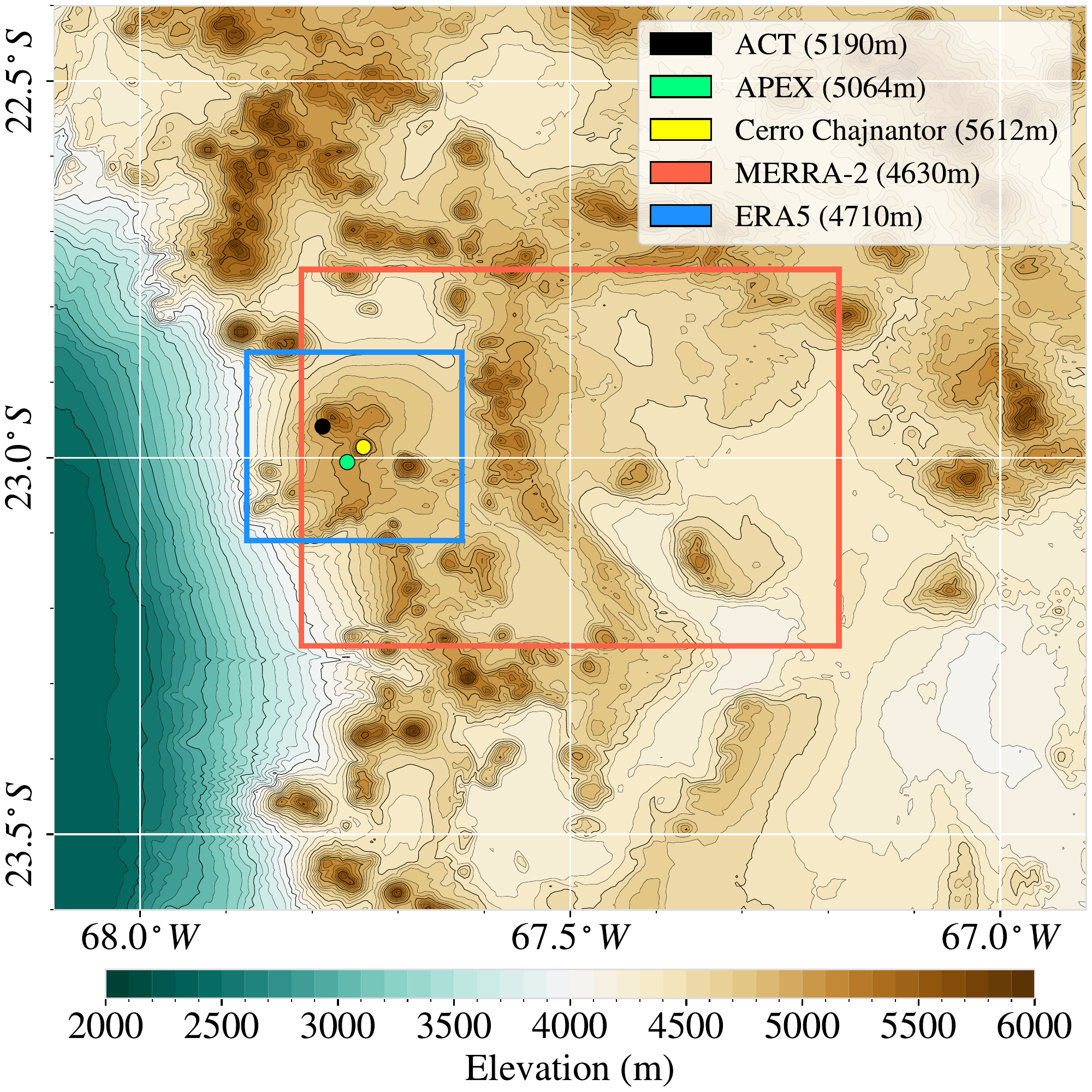}
\caption{Spatial footprints for each of the data sources considered in this paper, overlaid on a topographic map of the Chajnantor area. The elevations are obtained from the Shuttle Radar Topography Mission (\href{https://www2.jpl.nasa.gov/srtm/}{SRTM}) data set. ACT and APEX are fixed. APEX provides wind data at a resolution of one minute. The red box shows the $\sim 60$\,km square area averaged over by MERRA-2}
\label{fig:foot}
\end{figure}

Our analysis draws on a number of sources, including ACT \citep{thornton/etal:2016}, the ground-based weather station maintained by the Atacama Pathfinder EXperiment (APEX) collaboration  \citep[e.g.,][]{APEX:2009}, NASA's MERRA-2 database \citep{merra-2:2017}, the European Centre for Medium-Range Weather Forecasts (ECMWF) reanalysis \citep{ERA5}, the 
\citet{cortes/etal:2020} synthesis of the precipitable water vapor (PWV) in the Cerro Chajnantor region, and the UdeC-UCSC 183 GHz radiometer next to ACT \citep{bustos/etal:2014}. The location of each is shown in Figure~\ref{fig:foot}.

\begin{figure}
\includegraphics[width=1\columnwidth]{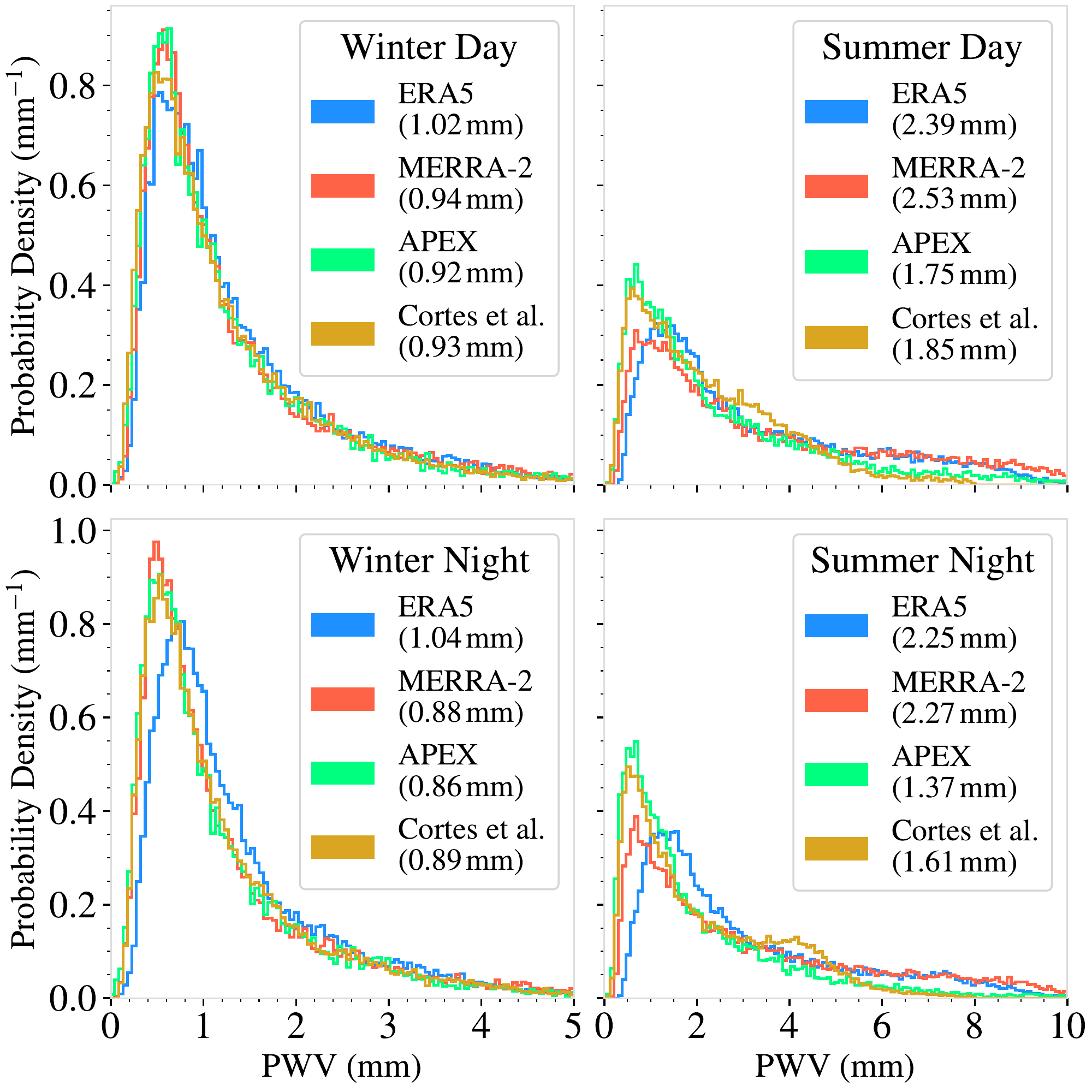}
\caption{Distributions of hour-averaged total atmospheric water vapor for several different sources in the Atacama Desert. The quantities in the legend for each histogram are the median PWV. In this paper, ``winter" describes the austral winter (May through October) and ``summer" describes the austral summer (November through April), while ``day" describes the hours between 11:00 and 23:00 UTC  and ``night" those between 23:00 and 11:00 UTC.}
\label{fig:pwv_hist}
\end{figure}

\begin{figure}
\includegraphics[width=1\columnwidth]{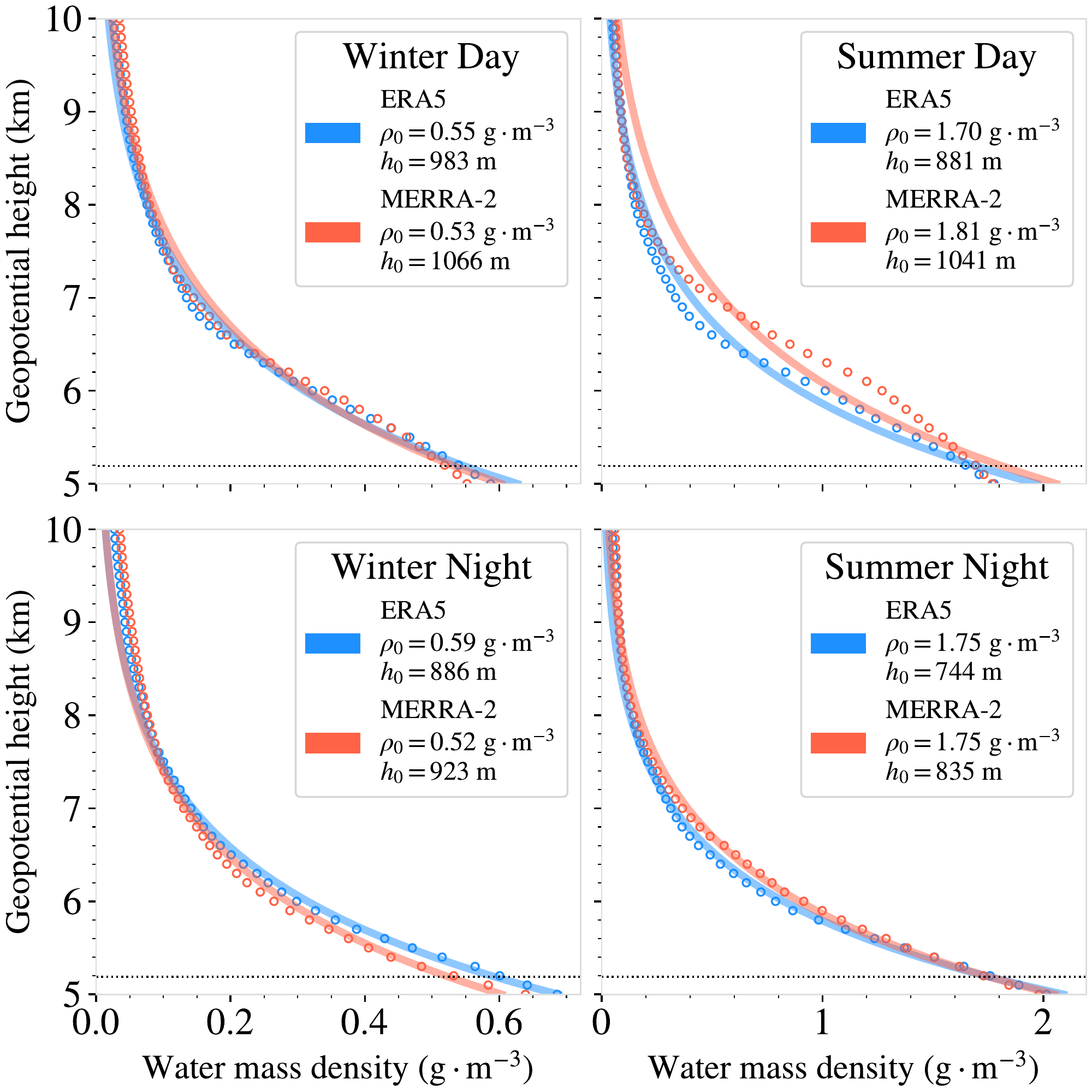}
\caption{Median water mass density profiles versus geopotential height, $h_g$, derived from atmospheric profile estimates for ERA5 and MERRA-2. The dotted line represents 5190 m, ACT's elevation. Fitting to Equation~\ref{eqn:theo_water_prof} yields a value for the half-height $h_0$ of around a kilometer. In contrast, the pressure scale height for the wet adiabatic atmosphere is roughly $5$\,km; thus the partial pressure of water decreases more quickly with height.}
\label{fig:profs}
\end{figure}

\subsection{ACT}
ACT is a 6-meter telescope that maps the CMB at millimeter wavelengths, located on Cerro Toco in the Atacama Desert at an elevation of 5190\,m. This paper considers around 15,000 hours of observation by ACT from between May 2017 and January 2021. For these data, ACT observed using three polarization-sensitive dichroic detector arrays (PA4, PA5, PA6), each having more than 1500 detectors with nominal frequencies of roughly 98 GHz (PA5, PA6), 150 GHz (PA4, PA5, PA6) and 220 GHz (PA4) \citep{henderson/etal:2016, li/etal:2016, ho/2017, choi/etal:2018,crowley/etal:2018}.
Each detector couples to a full-width half-maximum beam of 2.1, 1.4, and 1.0 
arcminutes for 98, 150 and 220 GHz respectively. The field of view for each array is a hexagon of corner-to-corner width 0.9$^\circ$. Although correlations between all pairs of arrays have been measured, in this paper we analyze each of the six array-band combinations independently of the others.

\subsection{APEX Weather Station}
The APEX weather station (referred to in this paper as APEX) is located on a 6-meter-tall freestanding structure located 50\,m west of the APEX telescope and approximately 5\,km south of ACT. Each minute it reports measurements of wind bearing, wind speed, and air temperature, as well as estimates of the total water column derived from the output of a 183 GHz radiometer. APEX is operated by the European Southern Observatory (ESO), and weather data is publicly available on the \href{http://archive.eso.org/wdb/wdb/eso/meteo\_apex/form}{ESO website}. APEX weather data in this paper comprises all available measurements between 2007 and 2020 (inclusive).

\subsection{MERRA-2}
The Modern-Era Retrospective analysis for Research and Applications, Version 2 (\href{https://gmao.gsfc.nasa.gov/reanalysis/MERRA-2/}{MERRA-2}) is a publicly-available database that combines satellite microwave observations to provide a comprehensive description of global weather. MERRA-2 data products are managed by the NASA Goddard Earth Sciences Data and Information Services Center, and are publicly available at the GES DISC website. This paper uses the~\href{https://disc.gsfc.nasa.gov/datasets/M2I3NVASM_5.12.4/summary}{M2I3NVASM data set}, that provides estimates for every 3-hour period from January 1980 to the present. 

The data set reports a set of variables that model the atmosphere at 72 roughly geometrically-spaced layers of geopotential height from around 4700 meters up to an altitude of 70\,km. The five variables that inform our analysis are the pressure, temperature, mass fraction of water vapor, and the northward and eastward components of the wind velocity. We used the data set centered at coordinates $67.5^\circ$W, $23^\circ$S for all three-hour periods between January 1980 and January 2021. MERRA-2 averages over a much larger area than ACT (longitudinal and latitudinal resolutions of $0.625^\circ$ and $0.5^\circ$, or around a 60\,km by 60\,km square) with a temporal resolution of three hours, which was resampled using a cubic spline to describe the atmosphere between a geopotential height of 5190\,m (the altitude of ACT) up to 20,000\,m at a vertical resolution of 100\,m and at a temporal resolution of one hour. Because of the variable topography, MERRA-2 measures atmospheric parameters at heights below that of ACT. In this paper, integrals of total atmospheric water vapor exclude the atmosphere below the height of the ACT site. 

\subsection{ERA5}

The ECMWF reanalysis, Version 5 (\href{https://www.ecmwf.int/en/forecasts/datasets/reanalysis-datasets/era5}{ERA5}) is a global reanalysis of weather data.
We use the \href{https://cds.climate.copernicus.eu/cdsapp#!/dataset/reanalysis-era5-pressure-levels}{1979-present pressure level data set} for which we consider dates from January 1980 to January 2021. Similarly to MERRA-2, ERA5 contains hourly estimates of weather parameters but at a finer spatial resolution ($0.25^\circ$ by $0.25^\circ$) and temporal resolution (hourly), though at a lower vertical resolution of 37 pressure levels. ERA5 was similarly resampled to the same vertical and temporal resolutions as MERRA-2.

\subsection{UdeC-UCSC Radiometer}

The UdeC-UCSC 183 GHz radiometer \citep{bustos/etal:2014} installed next to ACT has been measuring PWV since July 2018. It continuously records at 2-second resolution in a PWV range of 0.3--3.0 mm. This instrument was previously used by ESO for ALMA site testing and was refurbished at Universidad de Concepción in 2009. This paper considers all available data from July 2018 to January 2021. 

\subsection{Tipper Radiometers}

\citet{cortes/etal:2020} created a database of PWV measurements in the Chajnantor region using observations from two tipper radiometers that operated between 1997 and 2017. The locations of the radiometers changed over the course of their observations between the Chajnantor Plateau and the summit of Cerro Chajnantor; nevertheless, the results correlate well with APEX after slight adjustments for differences in elevation.

\subsection{Agreement of Weather Sources}

Using the pressure, temperature, and mass fraction of water vapor as determined by ERA5 and MERRA-2, we can obtain the mass density of water vapor as a function of height, and integrate over it to obtain an estimate for the total PWV.
The median and distribution of PWV measurements in the Chajnantor region as measured by APEX, ERA5, MERRA-2, and \citet{cortes/etal:2020} are shown in Figure~\ref{fig:pwv_hist}. We conclude that the PWV is generally consistent between different methods of measurement.

\citet{suen/fang/lubin:2014} compared ground-based water vapor radiometer measurements to satellite data at a number of different CMB sites around the world. They showed that the ground-based measurements are biased toward better observing conditions (lower PWV) because they are inoperable during bad weather. The higher PWV in the satellite data in the summer months as seen in Figure~\ref{fig:pwv_hist} is in qualitative agreement with this finding.

When considering measurements throughout the year, including times when the satellite data are available but the APEX data are not, ERA5 and MERRA-2 give a median PWV larger than APEX's by 0.43\,mm and 0.29\,mm respectively. When considering periods  
for which all three measurements are available, ERA5 and MERRA-2 are in rough agreement with APEX, overestimating the median PWV by 0.19\,mm and 0.04\,mm, respectively. The residual disagreement may be due to geographical and topographical differences between the sources.

\section{Atmospheric Emission}
\label{sec:emission}

Water vapor density decreases as a function of height, roughly following an exponential distribution defined by
\begin{equation}
\label{eqn:theo_water_prof}
    \langle \rho (h) \rangle = \rho_0 \exp \big [ -\log(2)\cdot(h_g - 5190\,\mathrm{m})\cdot h_0^{-1}\big ]
\end{equation}
where $h_g$ is the geopotential height. Figure~\ref{fig:profs} shows the median water vapor profiles for Chajnantor as measured by MERRA-2, and typical values for $\rho_0$ and $h_0$. Between 1980 and 2021, MERRA-2 estimates that 50\% of water vapor in the atmospheric column is within 2\,km of the ground 97\% of the time, and 90\% of water vapor is within 5\,km of the ground 93\% of the time. Typically, the half-height for water vapor $h_0$ is about a kilometer. 

Note that these profiles consist of averages over many years of data and do not accurately characterize variations on short time scales. In the physical atmosphere, turbulence introduces an inhomogenous and time varying distribution of water vapor, meaning that the actual line-of-sight profile of water vapor can deviate significantly from an exponential model. 
\newline

\subsection{Turbulent Distributions}
\citet{tatarski} showed that the mixing of passively distributed substances (like water vapor) in a turbulent velocity field evolves according to the same mechanism as the evolution of the velocity field, meaning that the distribution of water vapor in the atmosphere has the same spatial statistics as the distribution of velocity.

A useful approximation in modeling atmospheric water vapor is the Kolmogorov model \citep{kolmo}. It makes several simplifying assumptions about the time-dependent distribution of atmospheric velocities. Kolmogorov posits that an unconstrained, minimally viscous fluid (like the atmosphere) will be maximally turbulent and will thus have a velocity field with scale-invariant statistics. In three dimensions, water vapor is then distributed according to the spatial power spectrum
\begin{equation}
\label{eqn:kolmo}
    P(\mathbf{k}) \propto |\mathbf{k}|^{-11/3}.
\end{equation}

The Kolmogorov spectrum is not integrable, and thus physical turbulence cannot be scale-invariant for arbitrarily low $k$. Imposing a flat spectrum below some cutoff $k_{\mathrm{min}} = r_0^{-1}$, interpreted as corresponding to some maximum length scale $r_0$ on which the turbulence can still be said to be scale-invariant,\footnote{Typical values of $r_0$ for the Atacama are on the order of several hundred meters; see \citet{morris_thesis} and \citet{errard/etal:2015}.} 
leads to the adoption of an adjusted water vapor density spectrum 
\begin{equation}
\label{eqn:pspec}
    P_{\mathrm{adj}}(\mathbf{k}) \propto \big (r_0^{-2} + |\mathbf{k}|^2 \big )^{-11/6},
\end{equation}
which is normalized so that the total power is $\int_\mathbf{k} P_{\mathrm{adj}}(\mathbf{k}) d\mathbf{k} = 1$.
In order to consider spatial correlations in emission introduced by the turbulent atmosphere we use a model for the turbulent correlation function $D(r)$, which is obtained as the $d$-dimensional Fourier transform of the turbulent spectrum

\begin{equation}
    D(\mathbf{r}) \propto \mathcal{F}_d\big[P(\mathbf{r})\big](\mathbf{r}) = \int_{\mathbb{R}^d} P(|\mathbf{k}|) e^{-i\mathbf{k}\cdot\mathbf{r}}d^d\mathbf{k}. 
\end{equation}
For radially symmetric functions, the $d$-dimensional Fourier transform is also a radially symmetric function, and may be computed as 
\begin{equation}
    \mathcal{F}_d\big[P(k)\big](r) = (2\pi)^{d/2} r^{d/2-1}\mathcal{H}_{d/2-1}\big[k^{d/2-1}P(k)](r),
\end{equation}
where $\mathcal{H}$ is the Hankel transform.\footnote{The order-$\nu$ Hankel transform of $f(a)$ is given by the expression\begin{equation}
\mathcal{H}_\nu \big[f(a)\big](b) = \int_0^\infty f(a) J_\nu(ab) a da.
\end{equation}}
Plugging in Equation~\ref{eqn:pspec} with $d=3$ yields the isotropic correlation function 
\begin{equation}
\label{eqn:matern}
    D(r) = \frac{2^{2/3}}{\Gamma \big ( 1/3 \big )} \big (r / r_0 \big)^{1/3} K_{1/3} \big (r / r_0 \big),
\end{equation}
where $K_{1/3}$ is the modified Bessel function of the second kind of order $1/3$. This expression is normalized such that $D(0) = 1$.\footnote{Equation~\ref{eqn:matern} is Equation 1.33 in \citet{tatarski}. \citet{abramowitz/stegun:1970}
show that for small $z$, $K_{\nu}(z)\rightarrow 2^{\nu-1}\Gamma(\nu)z^{-\nu}$ (eq. 9.6.9).} The above correlation function was used to generate the turbulent atmosphere image in Figure~\ref{fig:peep}. 
We expect the relative strength of turbulent mass density fluctuations to follow the water vapor density and decrease as a function of height. We thus model the covariance of 
fluctuations as 
\begin{equation}
\label{eqn:turb_dens_cov}
    \big \langle \rho(\mathbf{r}) \rho(\mathbf{r}') \big \rangle = D \big (|\mathbf{r}-\mathbf{r}'| \big ) \big \langle \rho(\mathbf{r}) \big \rangle \big \langle \rho(\mathbf{r}') \big \rangle.
\end{equation}
where $\big \langle \rho(\mathbf{r}) \big \rangle$ is the expectation of water vapor density around a point $\mathbf{r}$ (Figure~\ref{fig:profs}). 

More difficult to model is the time-evolution of the distribution and the time-dependent correlations. Given that water vapor is passively distributed in a velocity field, we expect the velocity field to govern the time-evolution of the distribution. 
\citet{taylor} notes that turbulent velocities on small scales are small relative to the velocities on large scales. This means that turbulent distributions of water vapor will appear frozen on small scales, and that atmospheric features will be coherent as they move through a small angular aperture. This 
justifies a model referred to 
as the Kolmogorov-Taylor (KT) model, which translates some three-dimensional distribution of water vapor at some constant horizontal wind velocity $\vec{w}$ such that 
\begin{equation}
\label{eqn:KT}
    \rho(\vec{r},t) = \rho(\vec{r} + \vec{w} t, z).
\end{equation}
We use this approximation in the next section to outline a method of probing the mean angular motion of the atmosphere. 

\subsection{Modeling atmospheric emission}

CMB experiments employ arrays of detectors that convert incident electromagnetic radiation to a digitized signal. In practice, the signal includes various forms of contamination (thermal drifts, ground pickup, etc.), but fluctuations in atmospheric emission typically dominate the total spectrum of fluctuations from roughly $10^{-4}$\,Hz to 1--3\,Hz, above which the noise is approximately white and dominated by detector noise. 

\begin{figure}[htp]
\includegraphics[width=1\columnwidth]{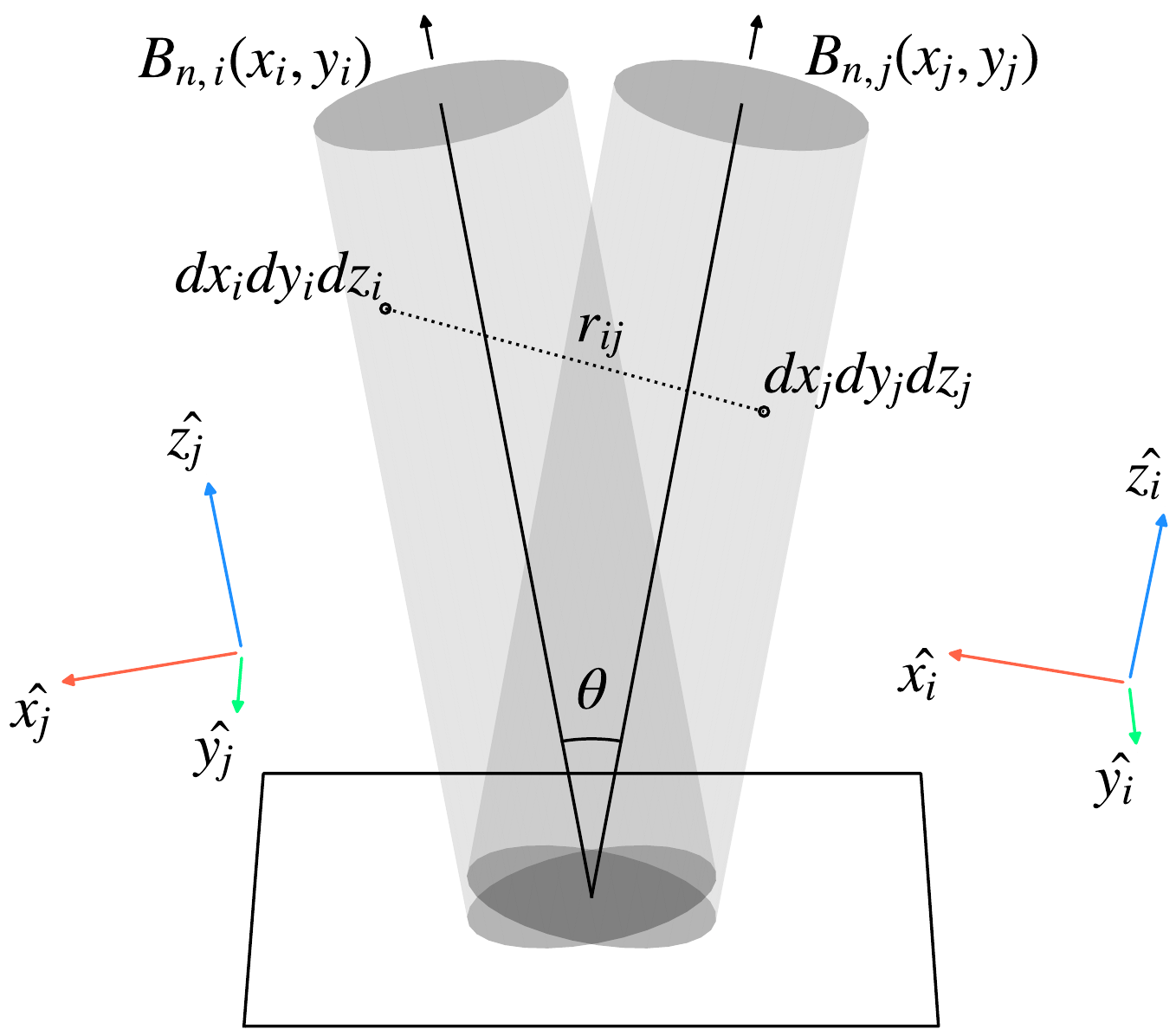}
\caption{A graphical representation of the beam-centered frames, for two beams separated by angle $\theta$. The beam separation here is exaggerated; this paper considers separations between $10^{-2}$ and $10^{1}$ degrees. The covariance of two detector temperatures is obtained by integrating over all pairs of volume elements within the geometries of the two beams, per Equation~\ref{eqn:six_integral_3}. The convention for this paper is that $\hat{y}$ always points toward the zenith.}
\label{fig:sep_frame}
\end{figure}

The power detected in a single mode of radiation with unit efficiency by a telescope observing an optically thin atmosphere in some direction $\mathbf{\hat{z}}$ at a given moment is given by
\begin{equation}
\label{eqn:1}
    P(t)=k_BT\Delta\nu =
    \frac{1}{2}\iiiint j_\nu(z,T)r(\nu)dA_n d\Omega d\nu dz,
\end{equation}
where $\Delta\nu$ is the bandwidth, $j_\nu(z,T)$ is the emission with units of Watt/m$^3$srHz as a function of distance in front of the telescope ($\int j_\nu dz$ is a surface brightness), $dA_n d\Omega=dA\cos\theta d\Omega$ is the differential element of the throughput or \'etendue, and $r(\nu)$ is the normalized instrument passband (see e.g. 
\citealt{condon/ransom:2016} for details). 
The coordinate system for this integral and what follows are  
``beam-centered" as shown in Figure~\ref{fig:sep_frame}; $x$ and $y$ are orthogonal coordinates corresponding to distances from the beam center, while $z$ is related to the height $h$ above the ground by $z = h \csc\epsilon$
where $\epsilon$ is the elevation (see Figure~\ref{fig:peep}).

We may simplify Equation~\ref{eqn:1} by considering a narrow frequency band around $\nu$, noting that in the Rayleigh-Jeans limit $j_\nu(z,T)=2k_BT_{\rm atm}\kappa(z,\nu)/\lambda^2$ where $\kappa$ is the absorption coefficient with units of inverse length and $T_{\rm atm}$ is the atmospheric temperature, and expressing the effective area as $A_e(\phi,\epsilon) = \lambda^2G(\phi,\epsilon)/4\pi=\lambda^2B_n(\phi,\epsilon)/\Omega_B$ where $G$ is the forward gain, $B_n$ is the normalized beam profile, $\phi$ and
$\epsilon$ are the azimuth and elevation, and $\Omega_B$ is the beam solid angle. For observations not too far from the optical axis Equation~\ref{eqn:1} reduces to
\begin{equation}
\label{eqn:det_temp}
    T =
   \frac{1}{\Omega_B} \iint \kappa(z,\nu)T_{\mathrm{atm}}(z,\nu)B_n(\phi,\epsilon) d\Omega dz.
\end{equation}
When pointing at the zenith in an isothermal and optically thin atmosphere the above reduces to
$\tau(\nu) T_\mathrm{atm}(\nu)$, where $\tau$ is the optical depth
\citep[e.g.,][]{condon/ransom:2016}.
We can relate $\kappa(z,\nu)$ to the water vapor density: considering only the water, $\kappa(z,\nu)=k_{H_{2}O}(\nu)\rho(z)/m_{H_{2}O}=\alpha_b(\nu)\rho(z)$ where $k_{H_{2}O}(\nu)$ is the molecular absorption coefficient (with units m$^2$), and $m_{H_{2}O}$ is the molecular mass. 
The proportionality constants for each frequency band, $\alpha_b(\nu)$, can be determined by calibrating to {\sl am} \citep{am:2018}. 

Because the illumination function of the primary resembles a tapered top hat, we approximate the beam as a tube of diameter $d_A=5.5$m up to the point where the diameter of the angular beam profile is greater than $d_A$ (9\,km at 90 GHz, 14\,km at 150 GHz, 19\,km at 220),\footnote{At an observing elevation of 45$^\circ$ these are well above most of the emission, but for a larger FWHM and field of view this would not be the case.} after which the profile grows at a constant angle equal to the FWHM. Another relevant scale is the height for which the beams from two adjacent feed horns no longer overlap; Figure~\ref{fig:act_beams} shows the angular size of the beams at different distances $z$ from the telescope. The separation between beams is $2^\prime$ for PA4 and $2.3^\prime$ for PA5 and PA6. In our simple model, beams of width $d_A$ no longer overlap at $z=8-10$\,km for $\epsilon=45^\circ$ corresponding to an altitude $h>5.6\,$km, well above the water vapor, again indicating that atmospheric fluctuations are correlated across an array.
In order to incorporate the turbulent statistics derived in the previous section, we want to reformulate Equation~\ref{eqn:det_temp} into the $(x,y,z)$ frame for the cylindrical beam approximation. The quantity $B_n(\phi,\epsilon)d\Omega dz/\Omega_B$ tells us to sum up $\kappa(z,\nu)T_\mathrm{atm}(z,\nu)$ over a region of space delineated by the beam, and so approximating the beam as a cylinder of diameter $d_A$, substituting in for 
$\kappa(z,\nu)$, and recognizing that the water density is time dependent yields
\begin{equation}
\label{eqn:beam_integral}
    T(t) =
   \alpha_b(\nu)\iiint \rho(x,y,z,t)T_{\mathrm{atm}}(z)
   B_n(x,y) dxdydz,
\end{equation}
where $\int B_n(x,y)dxdy=1$. An expanding beam may be accommodated by adding a $z$ dependence as in $B_n(x,y,z)$. Integrating over one line of sight in a small diameter cylinder beam is trivial. The above form becomes useful for examining the correlations between two nearby lines of sight. 

We adopt an exponential model for the mean water vapor density of the form
\begin{equation}
\label{eqn:water_prof}
    \langle \rho(z) \rangle = \rho(0) e^{-z / z_\rho},
\end{equation}
following the profiles shown in Figure~\ref{fig:profs}, where $z_\rho$ is around 1.5\,km at the zenith.\footnote{Note that $z_\rho$ is dependent on the pointing direction of the telescope.} We similarly adopt an exponential model for the temperature profile $T_\mathrm{atm}(z)$, which is well-approximated for heights below 10\,km by
\begin{equation}
\label{eqn:temp_prof}
    T_\mathrm{atm}(z) = T_\mathrm{atm}(0) e^{-z /z_T},
\end{equation}
where we ignore time-dependence or small-scale temperature fluctuations. In the Atacama and observing at the zenith, the average temperature at ground level is around $T_\mathrm{atm}(0) = 270$\,K, and $z_T$ is around 35\,km at the zenith.

\subsection{Angular atmospheric correlations}
\label{sec:angcorr}

\begin{figure}
\includegraphics[width=1\columnwidth]{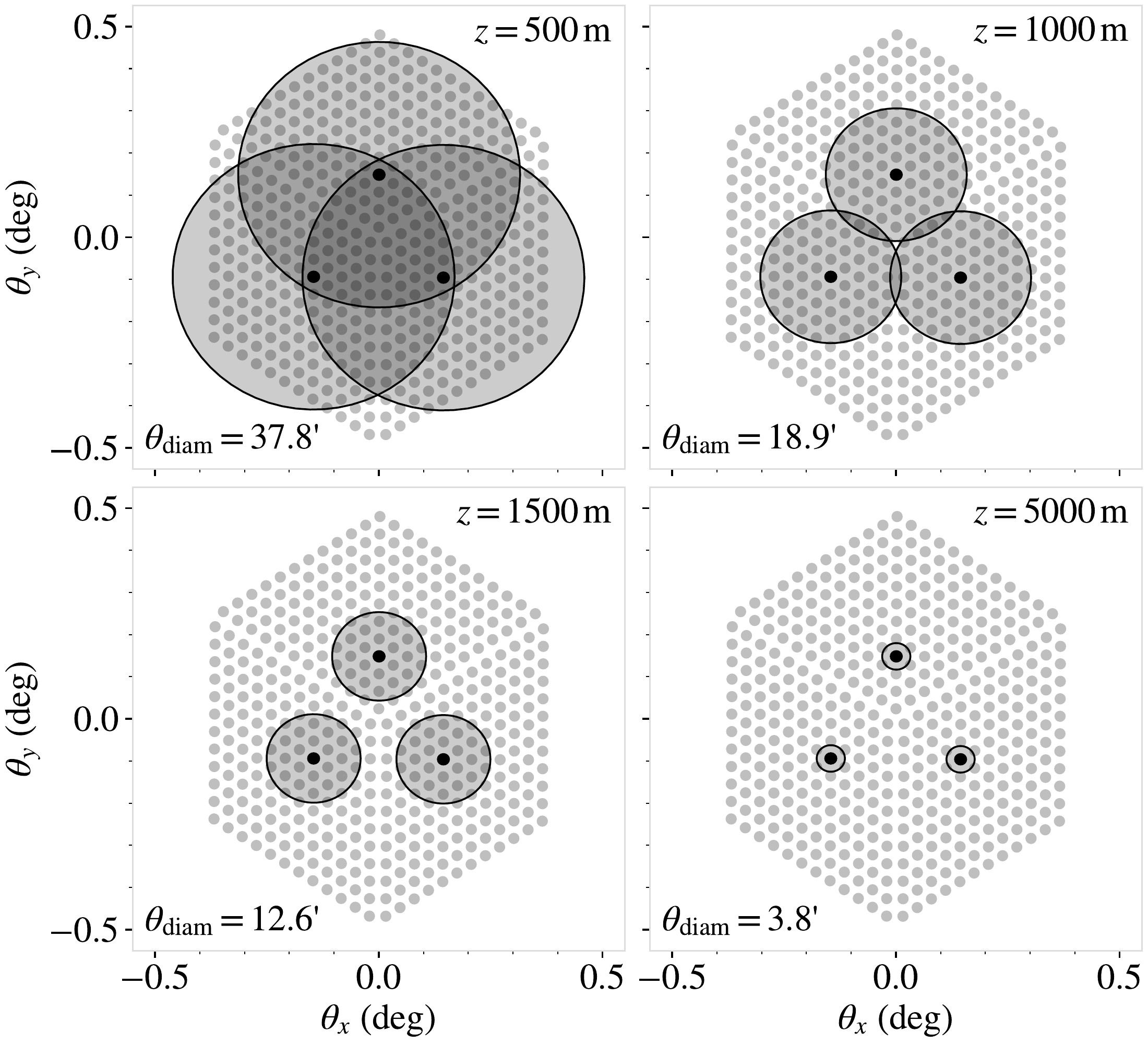}
\caption{The angular cross-section of ACT's PA6 array looking through the atmosphere. The large shaded circles represent the angular diameter $\theta_\mathrm{diam}$ of the beams for three detectors, at different atmospheric path lengths $z$. The dots indicate the beam centers and approximate far-field size of all detectors beams at 150\,GHz. At an elevation of $45^{\circ}$, the typical water vapor density halves every $1500$\,m along the beam. At a depth of $z=1500$\,m the beam diameter is $d_A=5.5$m and the width of the hexagonal array is 20\,m, and thus the beams heavily overlap, with their signals being highly correlated on scales of $\sim 1/4$ of the array.}
\label{fig:act_beams}
\end{figure}

\begin{figure}
\includegraphics[width=1\columnwidth]{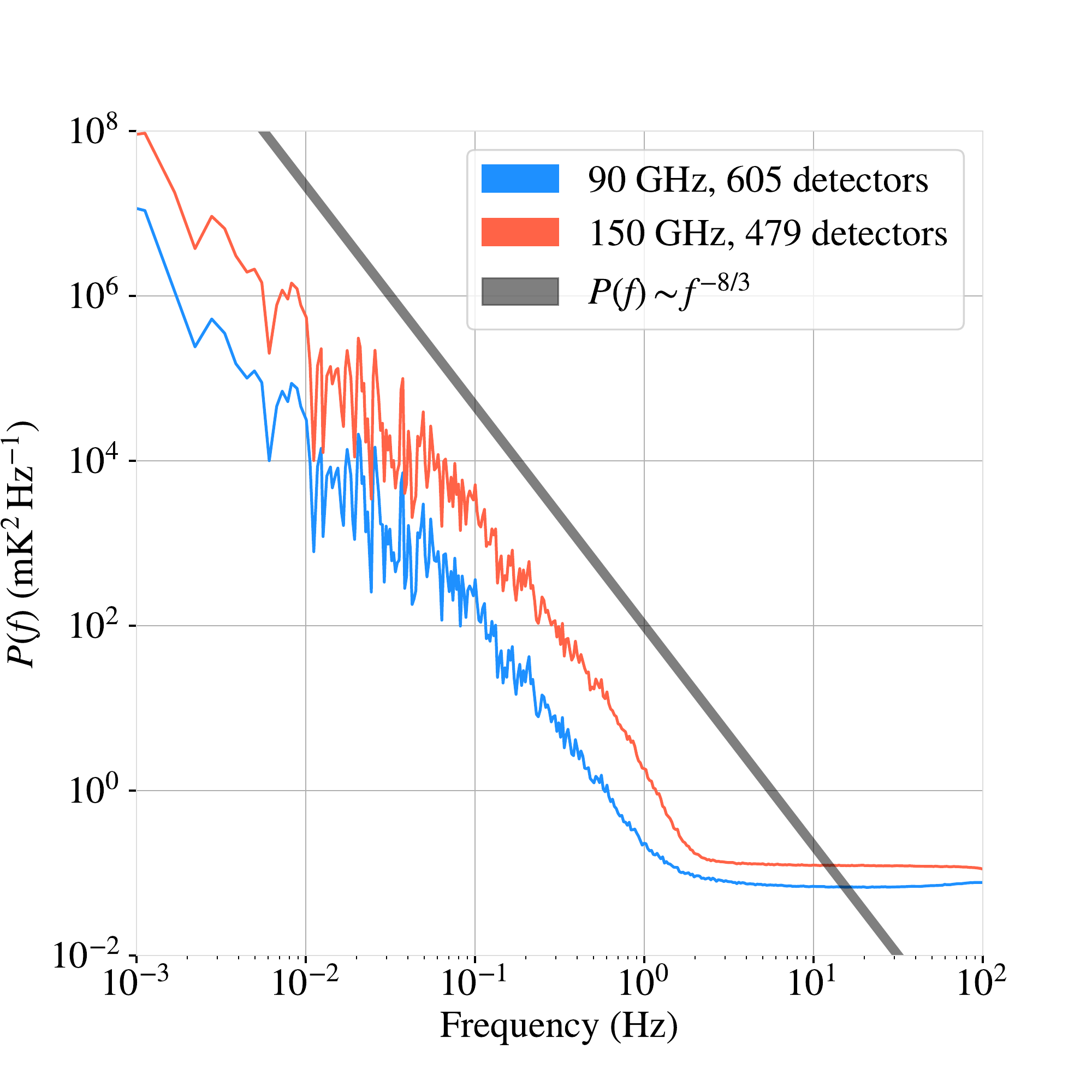}
\caption{The power spectrum of 33 minutes of ACT PA6 atmospheric stare data (when the telescope is stationary), averaged over several hundred detectors for the 90 GHz (blue) and 150 GHz (red) bands. The spectra of scanning data are qualitatively similar to stare data above the scanning frequency, which is typically below 0.1 Hz. The y-axis temperature fluctuations are relative to the CMB. Both bands follow a $-8/3$ spectrum (black) for low frequencies, which steeps near 1 Hz due to the finite-sized beam averaging over small-scale fluctuations, an effect explored in the appendix. At the time of observation, the PWV was 1.6\,mm and the wind speed at 1\,km was 8\,m/s, and thus a 5.5\,m large fluctuation crosses the beam at $\sim1.5$\,Hz. \citet{dunner/etal:2013} shows a similar plot as a function of PWV.} 
\label{fig:pspec}
\end{figure}

The angular covariance of the atmosphere can be computed by evaluating the expected product of Equation~\ref{eqn:beam_integral} between two detectors with respect to time. Consider two beams $i,j$ pointing with angular offsets $\vec{\theta}_i, \vec{\theta}_j$, separated by some angle $\theta = |\vec{\theta}_j - \vec{\theta}_i|$, where beam $i$ has an ACT-like cylindrical beam model $B_{n,i}$ for associated beam-centred coordinates $(x_i,y_i)$ (and analogously for beam $j$).\footnote{The convention for pointing angles is $\vec{\theta} = (\theta_x,\theta_y) = (x/z,y/z)$, and can be seen in Figure~\ref{fig:act_beams}.} Their covariance may be computed as 
\begin{widetext}
\begin{multline}
\label{eqn:six_integral_1}
    C(\theta) = \Big \langle  T(\vec{\theta}_i) \cdot T(\vec{\theta}_j) \Big \rangle = \alpha_b(\nu)^2 \Big \langle 
   \iiint \rho(x_i,y_i,z_i)T_\mathrm{atm}(z_i)
   B_{n,i}(x_i,y_i) dx_idy_idz_i \\ \times \iiint \rho(x_j,y_j,z_j)T_\mathrm{atm}(z_j)
   B_{n,j}(x_j,y_j) dx_jdy_jdz_j \Big \rangle 
\end{multline}
\begin{equation}
\label{eqn:six_integral_2}
    = \alpha_b(\nu)^2 \iiint \iiint \Big \langle 
   \rho(x_i,y_i,z_i) \rho(x_j,y_j,z_j) \Big \rangle T_\mathrm{atm}(z_i) T_\mathrm{atm}(z_j)
   B_{n,i}(x_i,y_i)
   B_{n,j}(x_j,y_j) dx_idy_idz_i dx_jdy_jdz_j 
\end{equation}
\begin{equation}
\label{eqn:six_integral_3}
    = \alpha_b(\nu)^2 \rho(0)^2 T_\mathrm{atm}(0)^2 \iiint \iiint D(r_{ij}) e^{-(z_i + z_j)(z_\rho^{-1}+z_{T}^{-1})}  B_{n,i}(x_i,y_i) B_{n,j}(x_j,y_j) dx_idy_idz_i dx_jdy_jdz_j, 
\end{equation}
\end{widetext}
where $r_{ij}$ is the distance between points $(x_i,y_i,z_i)$ and $(x_j,y_j,z_j)$ as shown in Figure~\ref{fig:sep_frame}, and where we include the turbulence density covariance model in Equation~\ref{eqn:turb_dens_cov} and the temperature profile model in Equation~\ref{eqn:temp_prof}. For the remainder of this paper we combine the density and temperature scaling parameters $z_\rho$ and $z_{T}$ into a single parameter $z_0 = (z_\rho^{-1} + z_{T}^{-1})^{-1}$, where the scaling parameter $z_0$ describes the decay of the strength of emission fluctuations.  Computing this integral (see Appendix~\ref{appen:a}) for very thin beams (such that $B_n(x,y) \to \delta(x)\delta(y)$) yields the expression
\begin{multline}
    C(\theta) = \frac{2^{1/6}\pi^{1/2}}{\Gamma\big(\frac{1}{3}\big)} A r_0 \int_0 ^\infty (z \theta r_0^{-1})^{5/6} \\ \times K_{5/6} \big [ z \theta r_0^{-1} \big ] e^{-2z/z_0} dz + B,
\end{multline}
where $A$ and $B$ are constants as defined in the appendix. When variables are highly correlated (as in the case of small angular separations), a more useful form of the covariance is the structure function, defined as
\begin{multline}
    \frac{1}{2} \Big \langle \big [ T(0) - T(\theta) \big ]^2 \Big \rangle = \Big \langle  T(0)^2 \Big \rangle - \Big \langle  T(0) T(\theta) \Big \rangle \\ = C(0) - C(\theta).
\end{multline}
When $r_0 \gg z\theta$ at all beam depths $z$ for which the emission contributes meaningfully and the beams do not heavily overlap,\footnote{A separation of $\theta = 1^\circ$ at $z = 3$\,km corresponds to $z\theta \approx 50$\,m, an order of magnitude less than the outer scale. Emission above a height of several kilometers is essentially negligible.} we may approximate the structure function as
\begin{equation}
    C(0) - C(\theta) \propto \theta ^{5/3},
\label{eqn:single_sum_5/3}
\end{equation}
which is derived in Appendix~\ref{appen:a}, and is observed as described in \citet{wollack/etal:1997}. This result also predicts a $-8/3$ power law, which is typically observed in the angular atmospheric power spectrum for ACT as seen in Figure~\ref{fig:pspec}. In the regime of small separations where the beams heavily overlap, the structure function more closely follows
\begin{equation}
\label{eqn:powlawtwo}
    C(0) - C(\theta) \propto \theta^2,
\end{equation}
as shown in Appendix~\ref{appen:a}. This effect is manifest in Figure~\ref{fig:pspec} where the spectrum steepens around 1\,Hz. Because ACT's beams heavily overlap for separations smaller than a degree and for heights from which we expect the majority atmospheric emission to be located, we use Equation~\ref{eqn:powlawtwo} to modify Equation~\ref{eqn:pl_sf} and take
\begin{eqnarray}
    C(0) - C(\theta) &\propto& \int_0 ^\infty (z \theta r_0^{-1})^2 e^{-2z/z_0} dz \nonumber\\
    &\approx& \sum_z \sigma^2_z (z \theta r_0^{-1})^2
\label{eqn:single_sum}
\end{eqnarray}
as a reasonably good approximation of the angular structure function as it appears to ACT for separations smaller than a degree. Note that we replace the exponential model of variance in water vapor density, $e^{-2z/z_0}$, with a more general function $\sigma^2_z$, that may be modeled using parameters obtained from ERA5 and MERRA-2. This particular form of the structure function has useful properties that we exploit below.

\section{Modeling Detector Correlations}
\label{sec:lags}

In this section we introduce the ``pair-lag" correlation and derive a physically-motivated
model for the time-evolving statistics of three-dimensional atmospheric emission relative to the array. In particular, this model allows for comparison of ERA5 and MERRA-2 data with ACT data by taking the structure function presented in Equation~\ref{eqn:single_sum} literally, and approximating the atmosphere as a discrete set of layers.

\subsection{The pair-lag correlation}

Consider an array of detectors with angular offsets $\vec{\theta}_i$ observing the atmosphere.
We first adopt a simplified model of brightness fluctuations, where the atmosphere consists of a single emissive layer $T(\vec{\theta})$ at some distance $z$ along the beam.

The atmosphere moves horizontally with 
linear velocity $\vec{w} = (w_\mathrm{east}(z),w_\mathrm{north}(z),0)$,\footnote{The quantity $w_\mathrm{east}$ describes an east-going wind, and thus comes from the west. In common usage, this is called a westerly wind.} which is variable as a function of distance along the beam. At distance $z$, this appears in the beam-centered frame as a two-dimensional angular velocity $\vec{\omega}(z) = (\vec{\omega}_x(z),\vec{\omega}_y(z))$, where
\begin{equation}
\begin{cases}
\omega_x(z) = -(w_\mathrm{east}(z)\cos\phi - w_\mathrm{north}(z)\sin\phi)z^{-1} \\
\omega_y(z) = -(w_\mathrm{east}(z)\sin\phi + w_\mathrm{north}(z)\cos\phi)z^{-1}\sin\epsilon.
\end{cases}
\end{equation}
with $\phi$ and $\epsilon)$ describing the azimuth and elevation. Two detectors observing the atmosphere with angular offsets $\vec{\theta_i},\vec{\theta_j}$ from the center of the array will observe a correlation in their observed brightness temperatures
\begin{equation}
    c_{ij} = \big \langle T(\vec{\theta}_i) T(\vec{\theta}_j) \big \rangle = C(z|\vec{\theta}_{ij}|r_0^{-1}),
\end{equation}
where $\vec{\theta}_{ij} = \vec{\theta}_j - \vec{\theta}_i$ is the two-dimensional pair orientation. Correlating between asynchronous samples with delay $\tau$, however, leads to an expression with a dependence on the angular velocity and the pair orientation,
\begin{equation}
    \label{eqn:ang_atm}
    c_{ij}(\tau) = \big \langle T(\vec{\theta}_i) T(\vec{\theta}_j + \vec{\omega}\tau) \big \rangle = C(z|\vec{\theta}_{ij} + \vec{\omega}\tau|r_0^{-1}).
\end{equation}
There is some unique delay $\tau_{ij}$ that maximizes the correlation of the two detectors $i,j$ by minimizing the argument $\vec{\theta}_{ij} + \vec{\omega}\tau$. This delay can be expressed as
\begin{equation}
\label{eqn:flat_lag}
\tau_{ij} = \underset{\tau}{\mathrm{argmax}}\big [ C(z|\vec{\theta}_{ij} + \vec{\omega}\tau |r_0^{-1})  \big ] = - |\vec{\omega}|^{-2}\vec{\theta}_{ij}\cdot\vec{\omega}.
\end{equation}
We refer to $\tau_{ij}$ as the ``pair-lag" of detectors $i$ and $j$, using the convention in \cite{morris_thesis}. Different versions of this quantity can be found in \cite{sayers/etal:2010} and \cite{robson/etal:2002}. The correlation between the two detectors is maximized when the detector separation is along the wind direction; in this case, the right-hand side of the equation gives the time for a fluctuation to get from one detector to another.
Note that Equation~\ref{eqn:flat_lag} does not offer a unique solution to $\vec{\omega}$ (which has two degrees of freedom) from the pair-lag of a single pair of detectors (which has only one). It requires at least two linearly-independent constraints, and thus at least three non-colinear detectors. Using three or more detectors, we can compute the pair-lag for each pair of detectors, and solve for the two-dimensional angular velocity $\vec{\omega}$ that best explains them.\footnote{The two components of $\vec{\omega}$ are defined with respect to the array, and assumes a flat focal plane so that the motion is uniform regardless of position on the array.}

\subsection{Approximating the Three-Dimensional Atmosphere}

We now present a model of the time-evolving three-dimensional atmosphere for use in Section~\ref{sec:apex_merra_compare} to allow comparison to data from ERA5 and MERRA-2.

We divide the atmosphere up into multiple layers, each still described by the same general model, only now with its own variance in the brightness temperature, $\sigma^2_z$, and wind velocity $\vec{w}(z)$. The model is no longer guaranteed to have an exact analytical solution, and instead becomes a maximization problem. The lag-dependent covariance of two beams is given by
\begin{equation}
    c_{ij}(\tau) = \sum_z \langle T_z(\vec{\theta}_i) T_z(\vec{\theta}_j + \vec{\omega}(z)\tau) \rangle,
\end{equation}
where for both computational and analytical feasibility we assume that emission that comes from different distances along the beam is uncorrelated, according to the single-sum model in Equation~\ref{eqn:single_sum}. Accounting for the turbulent scaling and covariance function, and ignoring correlations between layers, we then have 
\begin{eqnarray}
    c_{ij}(\tau) &=& \sum_z \sigma^2_z C(z |\vec{\theta}_{ij} + \vec{\omega}_z\tau |r_0^{-1})\nonumber\\
    &=& \sum_z \sigma^2_z \Big [ C(0) - (z |\vec{\theta}_{ij} + \vec{\omega}_z\tau |r_0^{-1})^2 \Big ]. 
\end{eqnarray}
%
This can be maximized as
\begin{eqnarray}
    \tau_{ij} &=& \underset{\tau}{\mathrm{argmax}} \sum_z \sigma^2_z \Big [ C(0) - (z |\vec{\theta}_{ij} + \vec{\omega}_z\tau |r_0^{-1})^2 \Big ]\nonumber\\
    &=& \underset{\tau}{\mathrm{argmin}}\sum_z z^{2} \sigma^2_z |\vec{\theta}_{ij} + \vec{\omega}_z\tau |^{2}.
\end{eqnarray}

Only the component of the angular separation parallel to the wind velocity matters, and so we write 
\begin{equation}
    \bar{\tau}_{ij} = \underset{\tau}{\mathrm{argmin}}\sum_z z^2 \sigma^2_z \big ( |\vec{\omega}_z|^{-1}\vec{\omega}_z\cdot\vec{\theta}_{ij} + |\vec{\omega}_z|\tau \big )^2,
\end{equation}
which we recognize as a weighted least-squares problem with respect to $\tau$. The solution is then 
\begin{equation}
    \bar{\tau}_{ij} =  - \frac{\sum_z z^2\sigma^2_z\vec{\omega}_z \cdot \vec{\theta}_{ij}}{\sum_z z^2\sigma^2_z|\vec{\omega}_z|^2}.
\end{equation}
We can define an aggregate angular velocity $\vec{\omega}_\mathrm{a}$ such that
\begin{equation}
\label{eqn:agg}
    |\vec{\omega}_\mathrm{a}|^{-2}\vec{\omega}_\mathrm{a} = \frac{\sum_z z^2\sigma^2_z\vec{\omega_z}}{\sum_z z^2\sigma^2_z|\vec{\omega}_z|^2}.
\end{equation}

This is the weighted harmonic mean of the variable angular velocity of atmospheric elements in the line-of-sight of the array. The aggregate angular velocity $\vec{\omega}_\mathrm{a}$ as defined by Equation~\ref{eqn:agg} averages over a dimension of space and thus cannot fully describe the motion of the three-dimensional atmosphere. However, we show in later sections that when applied to ACT it is computationally inexpensive to compute, provides a good effective characterization of fluctuations in atmospheric emission, and correlates well with the wind conditions at the ACT site as reported by external weather data sources (APEX, ERA5 and MERRA-2). The way $\vec{\omega}_\mathrm{a}$ averages over atmospheric depth in Equation~\ref{eqn:agg} is strongly dependent on the quantity $\sigma^2_z$, which describes the strength of the fluctuations in atmospheric emission as a function of the distance from the telescope. We model this quantity in Section~\ref{sec:apex_merra_compare}.

\subsection{Telescope Scanning}


Most ground-based CMB telescopes employ a constant-elevation, variable-azimuth scanning strategy in order to separate celestial signals from atmospheric contamination in subsequent map-making algorithms.\footnote{The results of this paper are generalizable to variable-elevation scanning strategies.} ACT, for instance, scans back and forth with an azimuthal speed of $1.5$ deg\,s$^{-1}$. This scanning motion introduces a relative atmospheric velocity that must be subtracted to determine the intrinsic atmospheric velocity,\footnote{At an elevation angle of $45^{\circ}$, the beam moves at around 25 m/s at an altitude of 1 km, comparable in magnitude to typical wind speeds.} as the aggregate angular velocities computed from the ACT data 
depend on the motion of the array. 
To account for the scanning motion, we transform the horizontal component of the aggregate angular velocity
\begin{equation}
    \vec{\omega}_{\mathrm{a,rel}} = 
    \begin{cases} 
    \omega_{\mathrm{a,rel},x} = \omega_{\mathrm{a},x} + \frac{d\phi(t)}{dt}\cos\epsilon(t) \\
    \omega_{\mathrm{a,rel},y} = \omega_{\mathrm{a},y} - \frac{d\epsilon(t)}{dt}, 
    \end{cases} 
    \label{eq:telmo}
\end{equation}
where $\vec{\omega}_{\mathrm{a,rel}}$ is the aggregate relative velocity of the atmosphere, and where $\phi(t)$ and $\epsilon(t)$ are the time-ordered azimuth and elevation of the array. This transformation allows us to switch between the array-relative and ground-relative frames for the atmospheric motion. 

Another effect of the scan is that in addition to imparting some apparent angular velocity to the atmosphere, it rotates the relative angle between the wind and scan directions as the telescope rotates through the width of its scan. We find that this effect is negligible for scan widths of less than 30$^\circ$. We address further effects of the scan in the next section, when we apply the pair-lag model to temperature data from atmospheric scans. 

\subsection{Atmospheric Velocities}

We define the aggregate angular wind velocity measured by ACT as
\begin{equation}
    \vec{u}_\mathrm{ACT} = 
    \begin{cases}
    u_\mathrm{east} = - \omega_{\mathrm{a},x}\cos\phi\csc\epsilon - \omega_{\mathrm{a},y}\sin\phi\csc^2\epsilon   \\
    u_\mathrm{north} = \omega_{\mathrm{a},x}\sin\phi\csc\epsilon - \omega_{\mathrm{a},y}\cos\phi\csc^2\epsilon . 
    \end{cases}
    \label{eqn:uvec}
\end{equation}
This definition allows us to compute the aggregate angular wind velocity from the aggregate angular motion given the elevation of the telescope, and has the intuitive interpretation of the speed at which the atmosphere appears to be moving to an observer looking at the zenith. It is also (in principle) independent of the angular elevation of the telescope, allowing us to compare the distribution of wind estimates from different telescope elevations.\footnote{Wind speeds on the ground in the Atacama Desert are typically on the order of 10 m/s, and tend to increase with altitude.} 

\section{Estimating Atmospheric Bulk Motion in ACT}
\label{sec:analysis}

We now estimate the aggregate angular motion from several years of ACT data. We apply the results derived in the previous section to build a database of wind velocity estimates. 

\subsection{Pre-processing and Detector Consolidation}

For this analysis, in contrast to mapmaking, ACT's raw time-ordered data\footnote{Time-ordered data (TOD) are stored in TOD files of roughly 10 min duration.} are down-sampled from 400\,Hz to 100\,Hz using an order-8 Chebyshev filter. Data were further filtered using an order-5 Butterworth filter so as to include only fluctuations between $10^{-1}$ and $10^{1}$ Hz. Faulty, dark, and otherwise undesirable detectors were excluded from the analysis.

One of the drawbacks of the pair-lag method is that it must perform $n_{pair} = n_{det}(n_{det} - 1)/2$ Fourier transforms in order to fully describe the correlations between $n$ detectors. In order to efficiently incorporate the entire array, detectors were grouped into 16 clusters and averaged together with their group, producing a number of ``consolidated detectors," each consisting of approximately 30--40 detectors. Clustering detectors is both more efficient and robust than considering individual detectors due to more desirable noise characteristics. 
The grouping washes out turbulent modes at scales below the grouping size, but these scales are not important for solving for the atmospheric motion; moreover, adjacent detectors are already highly correlated due to their heavily overlapping beams. 

\subsection{Sub-scan Division}

ACT scans with a variety of azimuthal widths, typically between $30^{\circ}$ and $80^{\circ}$. The small-angle scan approximation derived in the previous section can be exploited in practice even for wide-angle scans by dividing each total scan of constant azimuthal velocity to make smaller constituent sub-scans with sufficiently small half-widths. For ACT, each scan was divided into the maximum number of smaller sub-scans such that each one had an observation time of at least 8 seconds.\footnote{Because the duration of each scan is not perfectly divisible, smaller sub-scans were typically between 8 and 12 seconds long.} Because ACT scans with a constant azimuthal speed of $1.5$ deg\,s$^{-1}$, this corresponds to sub-scans of 
a half-width of $7.5^\circ$, for which the small-angle approximation is valid and for which the wind velocity does not appreciably rotate with respect to the array during the sub-scan. Dividing the scan into smaller sub-scans allows for a more accurate employment of the small-angle approximation, as well as higher-resolution measurements of time-dependent wind velocities. 

\subsection{Pair-lag computation}

We can compute the pair-lag of two detectors $i,j$ inexpensively as
\begin{equation}
\label{eqn:dft}
    \bar{\tau} = \bigg \langle \underset{\tau}{\mathrm{argmax}} \bigg [ \mathrm{DFT}^{-1} \Big [ \mathrm{DFT}\big[s^i_t\big]_f \cdot \overline{ \mathrm{DFT}\big[s^j_t\big]}_f \Big ]_\tau \bigg ] \bigg \rangle,
\end{equation}
where $s_i^t, s_j^t$ are their output signals, $\mathrm{DFT}[\ \cdot\ ]$ is the discrete Fourier transform and $\bar{[\ \cdot\ ]}$ is the complex conjugate. This approximation is valid for $f_{samp}^{-1} \ll |\bar{\tau}| \ll \Delta t$, where $f_{samp}$ is the sampling frequency and $\Delta t$ is the length of the sub-scan. Pair-lags were computed for each sub-scan for each unique pair of consolidated detectors using Equation~\ref{eqn:dft}. Pair-lags can sometimes deviate from those predicted by Equation~\ref{eqn:flat_lag} and return an unreasonable set of atmospheric parameters. This can be caused by singularities in the model (where the scanning motion and atmospheric motion nearly cancel out), or by non-atmospheric components of the signal, such as point sources or instrument glitches. In order to mitigate these effects, only non-zero pair-lags with a magnitude less than 2 seconds were considered.

\subsection{Fitting for the motion}

\begin{figure}[htp]
\includegraphics[width=1\columnwidth]{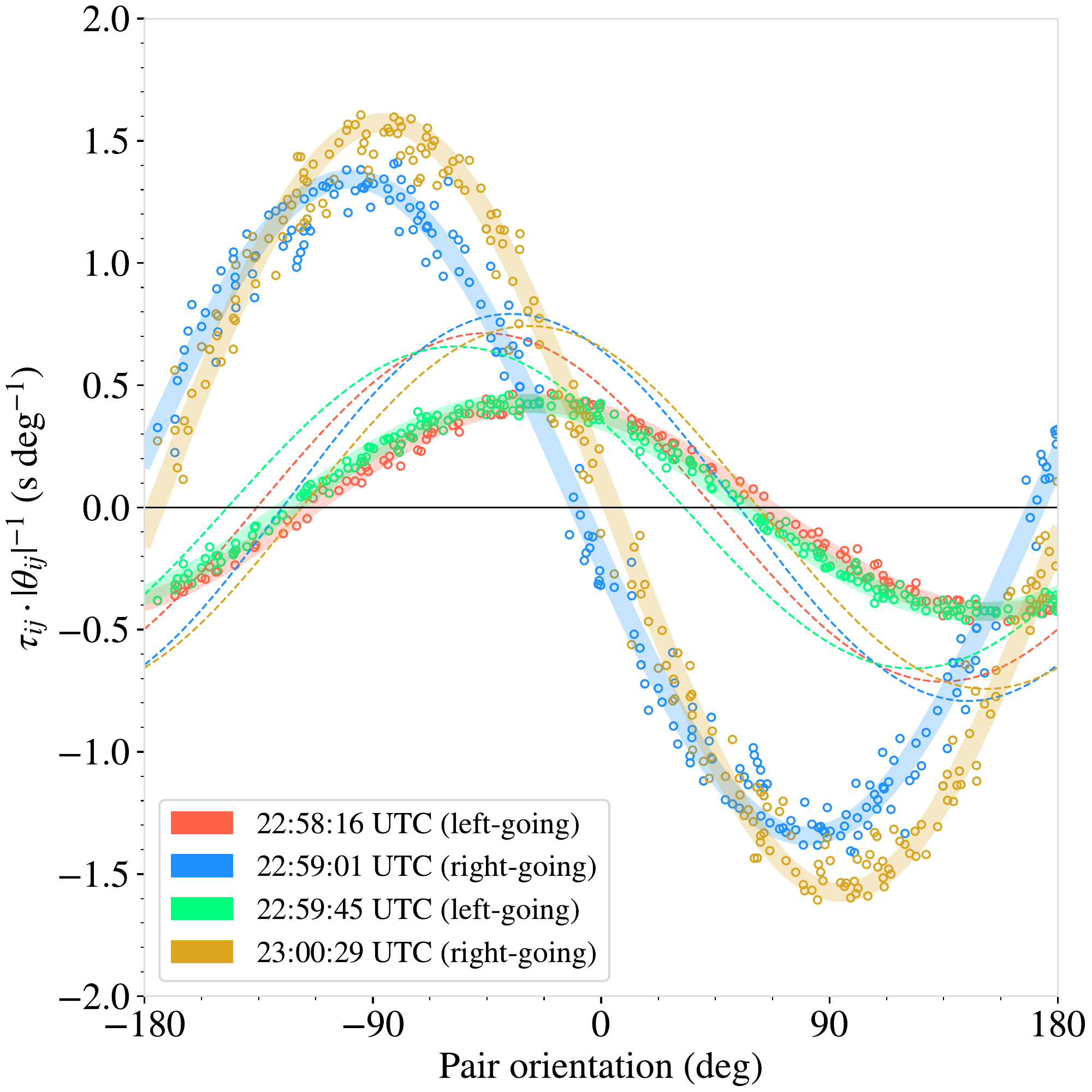}
\caption{A representation of the pair-lag model, applied to data from six consecutive scans of ACT; each scanning through 60$^\circ$ of azimuth over 40 seconds. The ratio of the pair-lag of two detectors to their separation roughly depends only on the angle of their orientation on the array. Here each point represents a pair of consolidated detectors. The motion of transient atmosphere across the array can be recovered from the fitted sine function (lightly shaded lines), where the direction is the phase of the function and the angular speed is the inverse of the amplitude. The contribution to the angular velocity by the scan can be removed, as shown by the dotted lines, so that each scan roughly agrees on the atmospheric velocity, which we attribute to the wind.}
\label{fig:pair-lag}
\end{figure}

Figure~\ref{fig:pair-lag} shows a plot of the pair-lags of 16 pairs of consolidated detectors divided by the distance between each consolidated pair versus the orientation of each pair on the array. The result is a sinusoidal relationship due to the dot product in Equation~\ref{eqn:flat_lag}. This indicates that over this time a constant angular speed projected onto the orientation of a pair of detectors is a good approximation. Note that the pair-lag divided by the separation on the array has units of inverse angular velocity. The magnitude of the measured velocity of the atmosphere across the array is given by the inverse of the amplitude of the sine wave. To find the net angular velocity of the wind relative to the array, we fit the pair-lags to the model in 
Equation~\ref{eqn:flat_lag}. This estimates the aggregate relative atmospheric velocity, which leads to a clear difference between left- and right-going scans (see Figure~\ref{fig:pair-lag}). After accounting for the telescope motion using Equation~\ref{eq:telmo}, we compute the components of the wind from the definitions of $u_\mathrm{east}$ and $u_\mathrm{north}$ in Equation~\ref{eqn:uvec}.

The ACT data correspond well to the linear pair-lag model and give generally consistent estimates of the atmospheric motion for all consolidated pairs. The high signal-to-noise
shown in Figure~\ref{fig:pair-lag} is typical for 80\% of the data. This behavior is consistent across several years of ACT observation, even when scans are further subdivided into sub-scans. 
Our results show that there are variations in the wind profile on the order of a few seconds, and that we can measure them consistently.

\vspace{0.2cm}
\begin{figure}[htp]
\includegraphics[width=1\columnwidth]{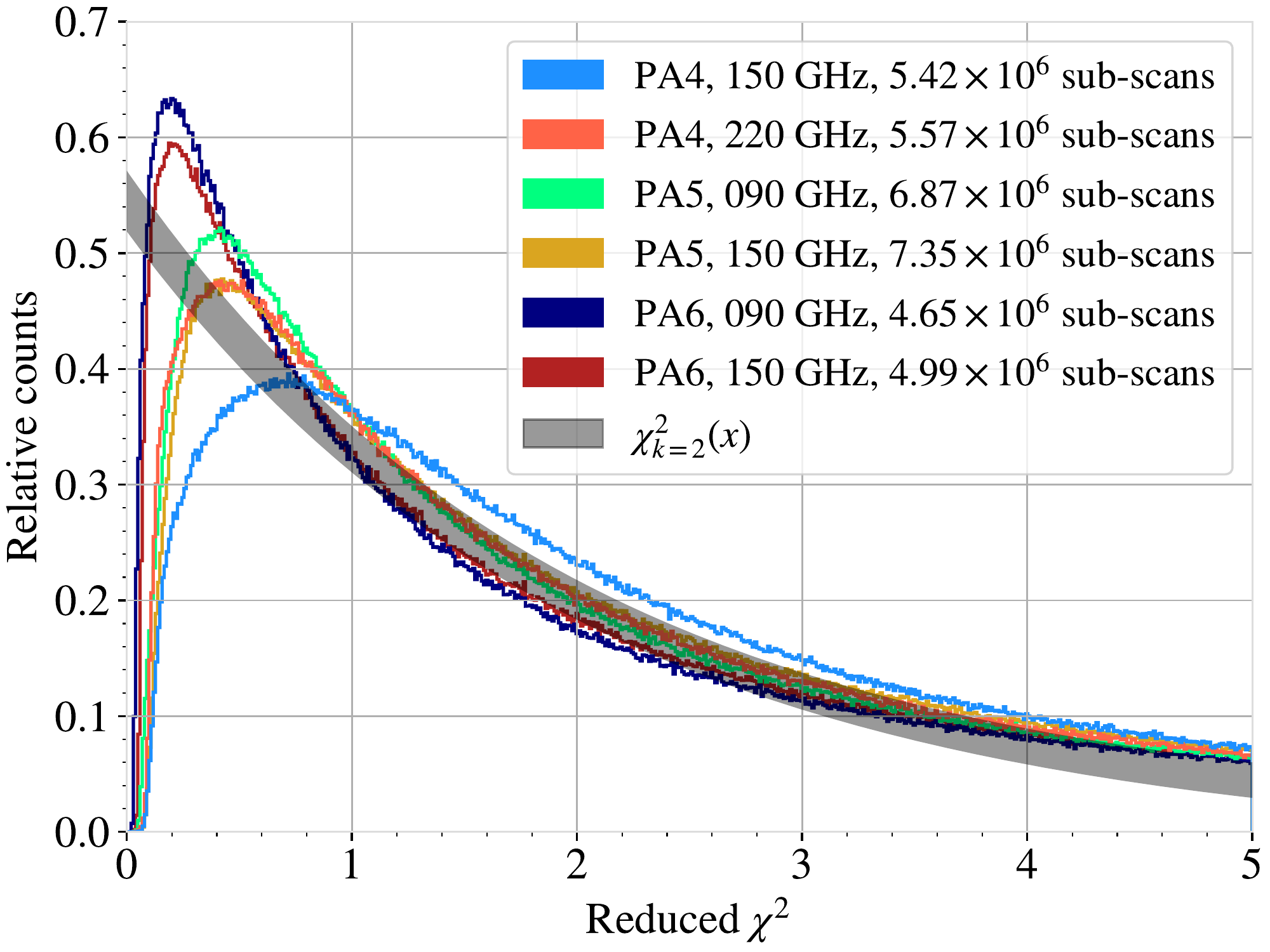}
\caption{The distribution of goodness of fits for wind estimates from each sub-scan, discriminated by array and observing band. Superimposed is the expected distribution for two degrees of freedom, normalized for $0 < \chi^2 < 5$. The goodness of fit is heavily dependent on the parameters themselves: a faster motion of the atmosphere with respect to the array is more easily and accurately detected by the model.}
\label{fig:chi2}
\end{figure}

\subsection{Processing Wind Velocities}

For each approximately 8--12 second sub-scan over four years, the analysis returns the two-dimensional angular wind velocity,
$\vec{u}_\mathrm{est}$ (Equation~\ref{eqn:uvec}), along with a chi-squared goodness of fit parameter, $\chi^2$, and the time $t$, azimuth $\phi$, and elevation $\epsilon$ of the center of the sub-scan.\footnote{Due to the segregation of array-band combinations, each unique point in time then has between 0 and 6 estimations of the wind velocity corresponding to the two frequencies in the three arrays.} 
The derived wind speeds vary in quality due to myriad factors, the most prominent being the effects of irregular non-atmospheric signals in the data. We excluded wind estimates with a speed greater than $5$ deg\,s$^{-1}$, as well as estimates for which the parameter estimator did not converge. These cases corresponded to 20\% of the data. 

To compute the $\chi^2$, we take the variance of each pair of consolidated detectors to be equal to 
the median variance from the sub-scans after removing the best-fit two-parameter model for each array-band separately. Figure~\ref{fig:chi2} shows the distribution of the goodness of fit for each array-band combination. We kept estimates with $\chi^2<5$, which corresponds to 75\% of all remaining estimates. Most of time the model gives a reasonable fit to the data, and it is apparent when it does not. 

The fit results are irregularly sampled due to breaks in data acquisition (for example, for calibration or planet mapping). For each ten-second bin centered at time $t$, the smoothed wind estimation is given by the weighted average of the raw wind estimates from all sub-scans and all arrays as
\begin{equation}
\vec{u}_\mathrm{ACT}(t) = \frac{\sum_i m_i(t) \vec{u}_{\mathrm{est},i}}{\sum_i m_i(t)}
\end{equation}
with the weights given by 
\begin{equation}
\label{eqn:weights}
    m_i(t) = \frac{1}{\chi^2_i} \cdot  \exp{\Big [ -\frac{(t-t_{i})^2}{2\sigma_t^2} \Big ]},
\end{equation}
where $\vec{u}_{\mathrm{est},i}$, $\chi^2_i$ and $t_{i}$ are the wind estimate, the goodness of fit and the time of the sample for the $i$-th sub-scan. For this paper, we choose $\sigma_t = 10$\,s. Bins more than a minute away from any estimate are deemed not to have an estimate. 

We find that there is typically a slight difference in the distribution of wind estimates obtained from left-going and right-going scans, which most likely arises from the approximation about the behavior of the pair-lags as the motion of the atmosphere interacts with the angular motion of the array. We mitigate this by adjusting the weights in Equation~\ref{eqn:weights} such that for any time $t$, exactly half the weight comes from each scan direction. 

%
%
\section{Analysis of ACT-derived atmospheric motion}

\begin{figure*}
\epsscale{1.1} 
\plotone{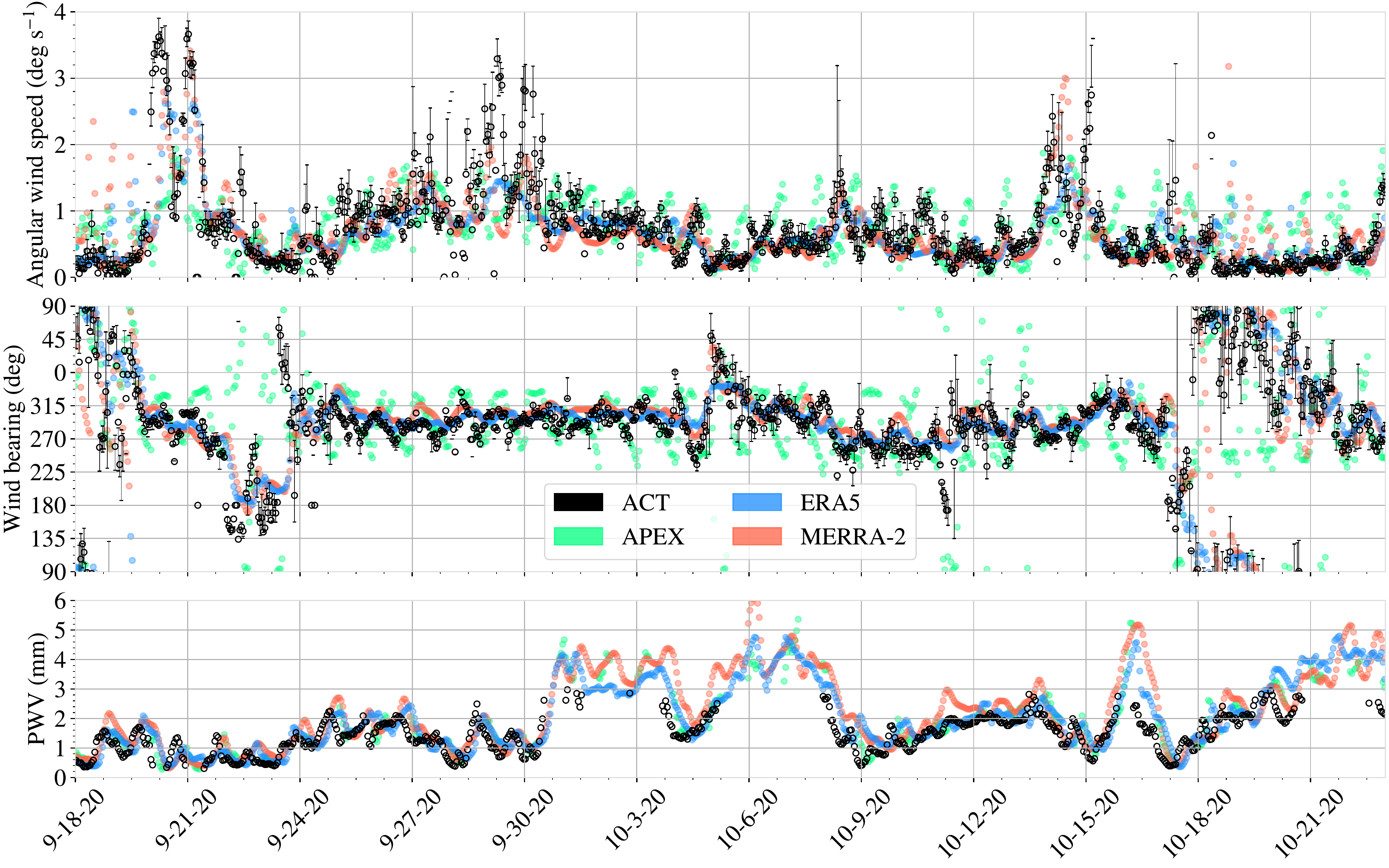}
\caption{Time-ordered hourly binned estimates for angular wind speed, wind bearing, and PWV from each of ACT, APEX, MERRA-2, and ERA5. PWV measurements from the ACT site (bottom panel, black) are from from the UdeC-UCSC radiometer that reports PWV in the 0.3--3.0 mm range. ACT-derived wind estimates (black), superimposed on the aggregate angular wind velocities for APEX (green), MERRA-2 (red), and ERA5 (blue) derived as described in the text, from September to October of 2020. The error bars for ACT-estimated winds represent the middle two quartiles (25\% to 75\%) of the distribution of each quantity in each bin. Some deviation in these various measures is expected given the different analyses and types of measurement.}
\label{fig:time_comp}
\end{figure*}

\begin{figure*}
\epsscale{1.1} 
\plotone{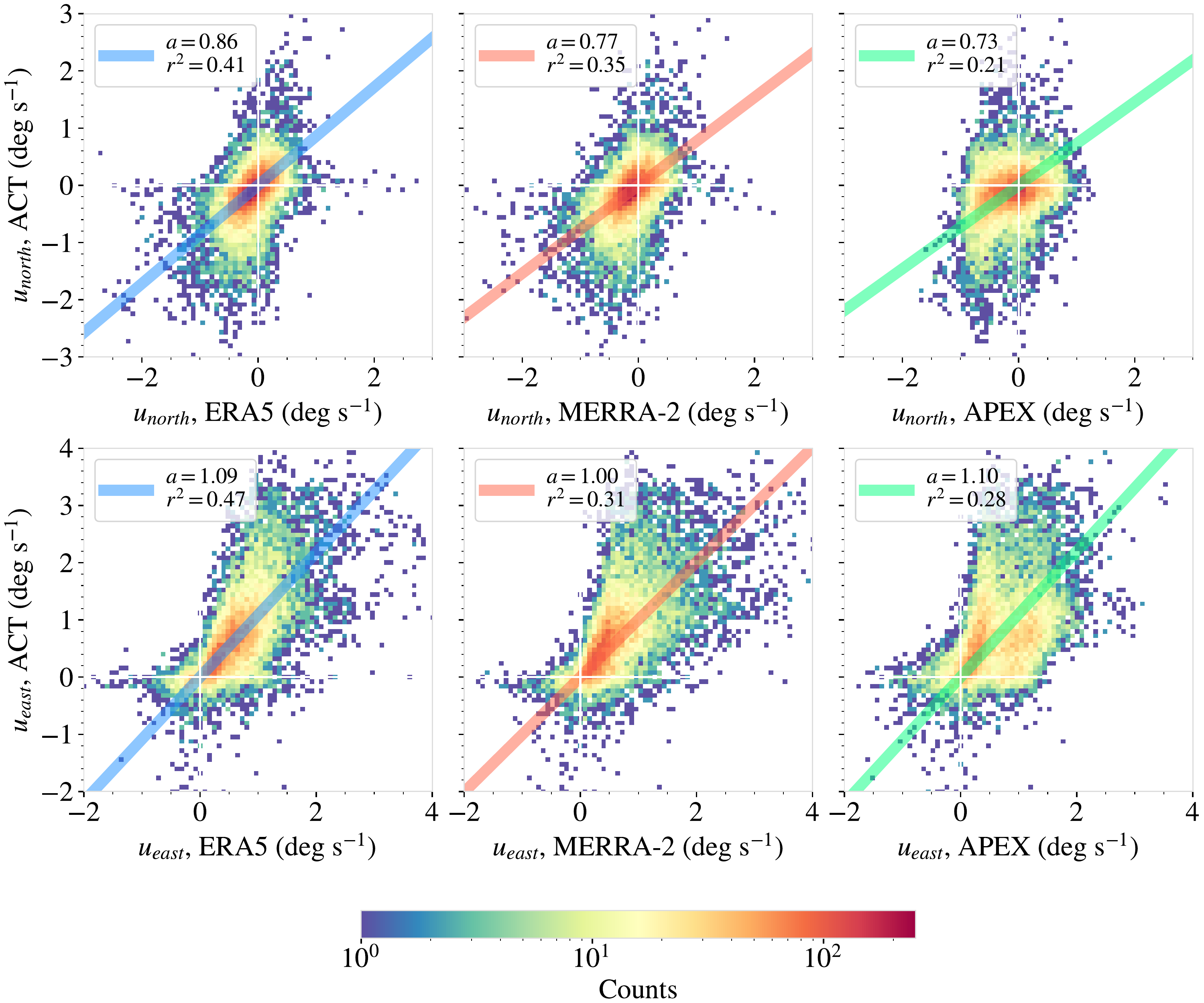}
\caption{Correlations of different weather sources with ACT-derived angular wind estimates, each with the weighted best-fit linear determination (weights are given by the confidence in ACT estimates). Each plot represents the roughly 12.5\,khrs of data for which ACT, ERA5, MERRA-2, and APEX data were available between May 2017 and January 2021 (38\% of the time). The slope of the best weighted linear fit between sources is typically within 10\% of unity for all sources. Non-linear factors are apparent in the correlation; in particular, the model tends to underestimate ACT-derived wind speeds in the summer, and overestimate them in the winter as shown in Figure~\ref{fig:hist_comp}.}
\label{fig:corr_comp}
\end{figure*}

\begin{figure}

\includegraphics[width=1\columnwidth]{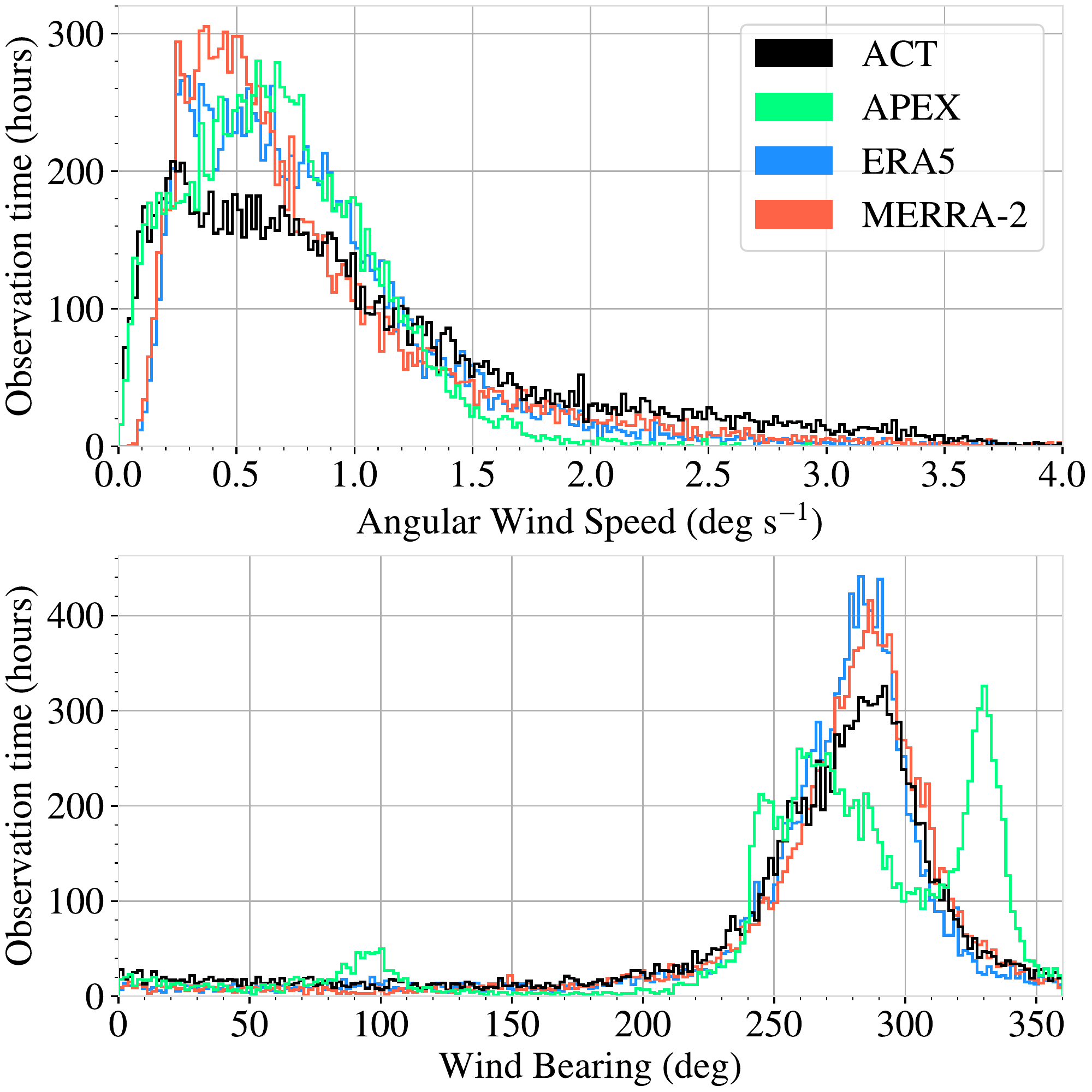}
\caption{A histogram of the median aggregate angular wind speed and median bearing of each data set for all hour-long periods in which all four data sets were available. There is generally good agreement between these different measures.}
\label{fig:hist_comp}
\end{figure}

\begin{figure}
\includegraphics[width=1\columnwidth]{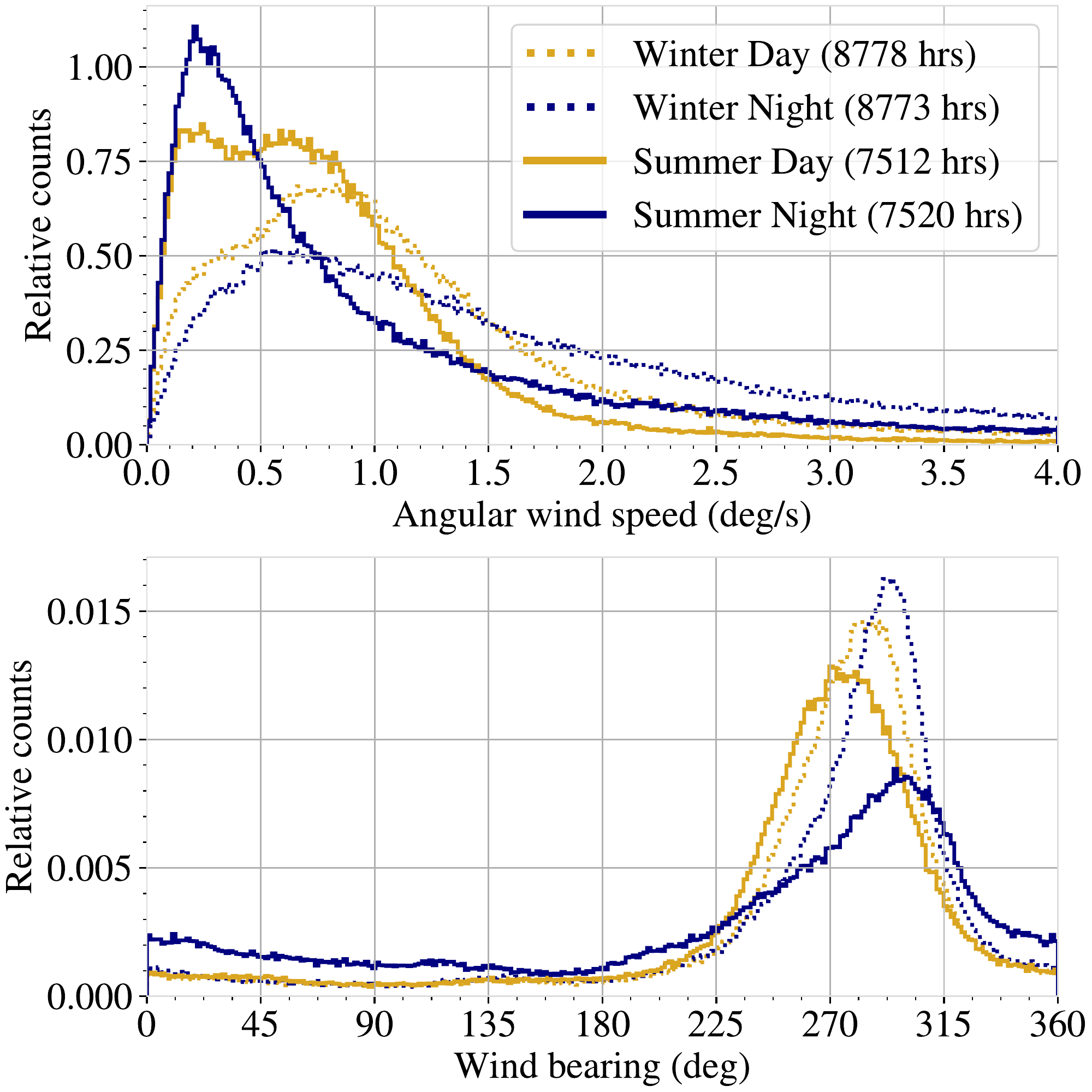}
\caption{The distribution of ACT-derived smoothed angular wind velocities with a one minute time scale, discriminated and normalized by season and time of day. Winter months typically have higher wind speeds than summer months.}
\label{eqn:act_wind_seg}
\end{figure}

\subsection{Comparison to Weather Data}
\label{sec:apex_merra_compare}

Comparing wind data from ACT, APEX, ERA5, and MERRA-2 must be done with care because each source measures a fundamentally different aspect of the atmosphere: ACT describes the aggregate angular motion of atmosphere fluctuations, APEX measures the linear wind velocity near the ground, and ERA5 and MERRA-2 measure the atmosphere at a series of discrete heights. Nevertheless, we can investigate the explanatory power and limitations of each data set on the other. For all sources, converting to a form directly comparable to ACT-derived estimate requires the assumption of some atmospheric model. 

In the case of APEX, which provides the physical wind speed and direction, the wind vector $\vec{w}_\mathrm{APEX}$ must be divided by some scale height $h_\mathrm{APEX}$ in order to obtain an angular wind velocity
\begin{equation}
\vec{u}_\mathrm{APEX} = \vec{w}_\mathrm{APEX}\,h_\mathrm{APEX}^{-1}.
\end{equation}
This scale height was determined 
by minimizing the median difference in the hour-averaged angular wind velocity estimates for ACT and APEX during all hour-long periods for which both ACT and APEX estimates were available (approximately 10\,khrs). This yields a scale height of $h_{\rm APEX} \approx 400$\,m. Note that this quantity does not necessarily represent the effective height of atmospheric turbulence: in the Atacama Desert, wind speed generally increases with height which biases the inference toward lower scale heights. 
However, this result approximately agrees with the effective height of turbulence in phase fluctuations\footnote{Phase fluctuations arise from variations in the index of refraction as opposed to water vapor density.} found by \citet{robson/etal:2002}, who assume a constant wind profile and find a scale height generally on the order of 500\,m. (See also \citealt{perez/rivera/nyman:2005}.)

Atmospheric reanalysis data sets like ERA5 and MERRA-2 allow us to model the aggregate angular motion as derived in Section~\ref{sec:lags}. We use this model to compute the aggregate angular wind velocity for ERA5 and MERRA-2 using the formulation derived in Section~\ref{sec:lags} as
\begin{equation}
\frac{\vec{u}_\mathrm{a}}{|\vec{u}_\mathrm{a}|^{2}} = \frac{\sum_h h^2 \sigma^2_h \vec{u}_h }{ \sum_h h^2 \sigma^2_h |\vec{u}_h|^{-2}},
\end{equation}
where $\vec{u}_h = \vec{w}_h \, h^{-1}$ is the angular wind vector at height $h$ based on the physical velocity reported by each data set. This necessitates a statistical model of the relative strength of fluctuations in emission as a function of height, $\sigma^2_h$. \citet{church:1995} and \citet{errard/etal:2015} approximate the variance of the fluctuations as being proportional to the water vapor mass density and the physical atmospheric temperature. In Section~\ref{sec:emission} we modeled the emission profile as an exponential function, the product of exponential profiles of water density and temperature. ERA5 and MERRA-2 allow us to be more specific, however, providing the explicit
water density and temperature profiles. We thus model  
\begin{equation}
\label{eqn:var_prof}
    \sigma^2_h \propto \big ( \rho_\mathrm{RA}(h) T_\mathrm{atm,RA}(h) \big )^2 .
\end{equation}
Here $\rho_\mathrm{RA}(h)$ and $T_\mathrm{atm,RA}(h)$ are the reanalysis profiles water density and temperature as a function of height that are provided by ERA5 and MERRA-2 at hourly increments. 

The model for $\sigma^2_h$ derived using data from ERA5 and MERRA-2 has a typical half-height of around $500$\,m, which is roughly half the half-height of total water vapor density ($h_0 \sim 1000$\,m). This is in rough agreement with fitted $h = 400$\,m for APEX; the slight discrepancy may be explained by the fact that wind speeds typically increase as a function of height, which is apparent in ERA5 and MERRA-2, and in other studies of Atacama weather \citep[e.g.,][]{masciadri/etal:2013}. 

We note that the half-height of $\sigma_h^2$ describes the variance in emission. Similarly, the angular speed determined from the pair-lag is based on that variance. For an exponential distribution of water vapor (Figure~\ref{fig:profs}) with half-height $h_0$, the half-height of the variance is $h_0/2$. Thus the effective half-height of the modeled emission is consistent with the measured distribution of water vapor. 


\subsection{Agreement with Weather Data}

Figure~\ref{fig:time_comp} shows a time-ordered comparison for the four sources over two months in the austral spring of 2020. ACT-derived wind data are effective at detecting changing wind directions in the upper atmosphere, and corresponds more closely to ERA5 and MERRA-2 than APEX. The four sources of wind data can sometimes differ substantially in their prediction of the aggregate angular wind, most likely due to the inability of the emission profile model to capture variations in the characteristics of the atmosphere on short timescales. ERA5 and MERRA-2 also average over larger spatial footprints, whereas ACT averages over the projection of a small focal plane through the atmosphere.

Figure~\ref{fig:corr_comp} shows the correlations of northward and eastward angular wind speeds for each of ERA5, MERRA-2 and APEX with ACT. In conjunction with Figure~\ref{fig:time_comp}, it shows that while there is a clear relationship between the weather sources and ACT, they do not predict the wind velocity from ACT with a consistent slope. The predictive capacity of the weather sources on ACT data might benefit from added degrees of freedom in the scaling between the two, but this is beyond the scope of this paper. 

Figure~\ref{fig:hist_comp} shows the distributions of angular wind speed and direction for each source. The three-pronged distribution of wind bearings from APEX is caused by a large diurnal variation in the wind; such variations in the wind are most pronounced near the ground. ACT, ERA5, and MERRA-2 directions are determined largely by the more consistent upper atmosphere. 

Figure~\ref{eqn:act_wind_seg} uses the ACT-derived properties of the wind over an observing period of 4 years. Although the bearing is almost always westerly, there is a significant seasonal variation in the distribution of wind speeds.

We conclude that external weather sources such as APEX, ERA5, and MERRA-2 describe the atmosphere as it appears to millimeter-wave telescopes, at least when averaged over timescales of an hour. We can also see roughly the same scaling of angular velocities in both figures, which lends credence to both the approximation derived in Section~\ref{sec:lags}, as well as the model that fluctuations scale with the total density. However, we find that the best source of data about the atmospheric motion as it appears to ACT is, likely, ACT itself via the pair-lag model, as it can attain a finer spatial and temporal resolution than ERA5 and MERRA-2.
Accurately estimating changes in velocity is essential to understanding the characteristics of atmospheric fluctuations, as we show in the next section. Ultimately, the correctness and usefulness of the pair-lag model will be ascertained by how well it can be used to mitigate the effect of atmospheric noise in the data analysis but that is beyond the scope of this paper.   

\subsection{Effects of bulk atmospheric motion on time-ordered spectra}

\begin{figure}[th]
\includegraphics[width=1\columnwidth]{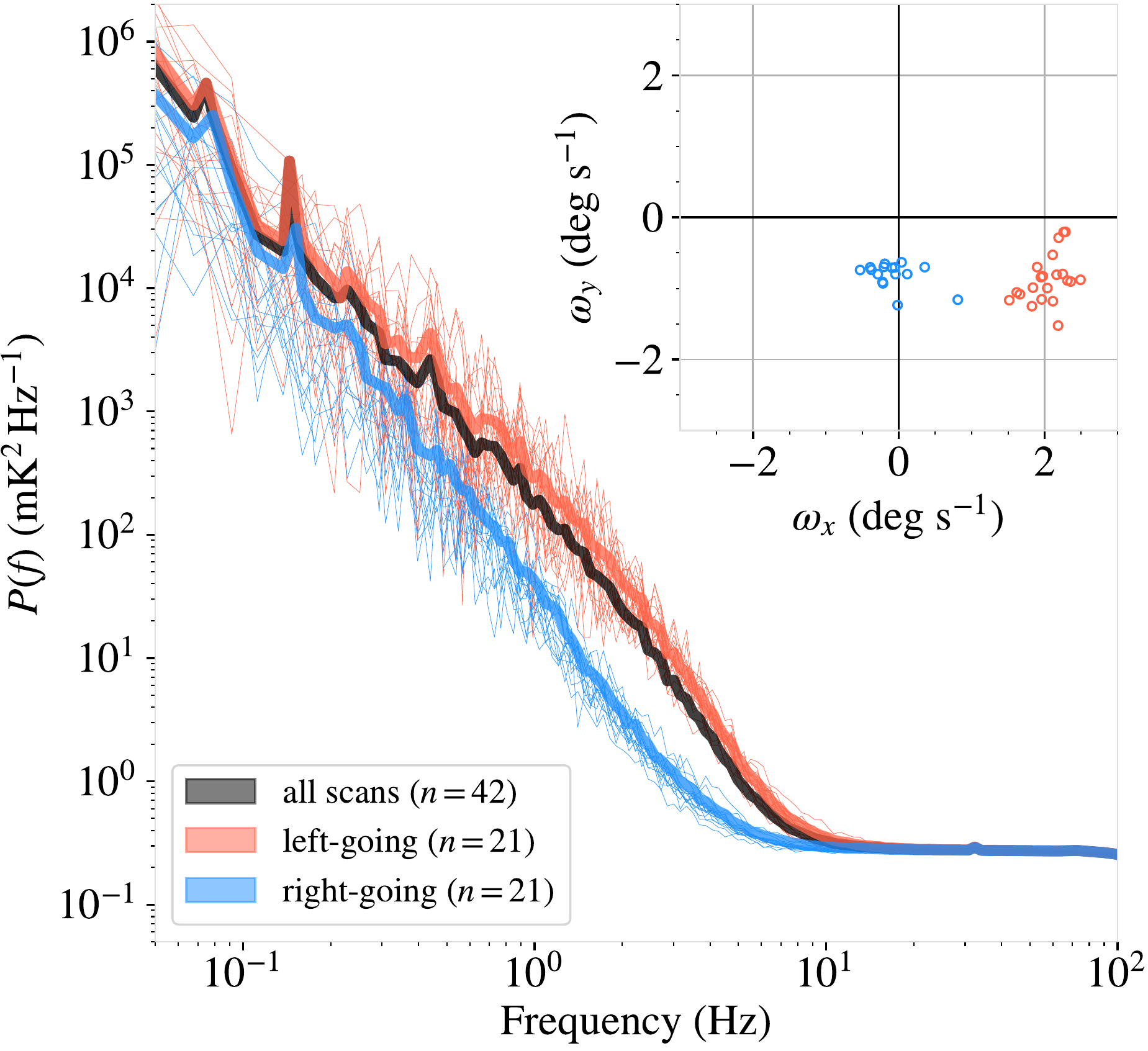}
\caption{The power spectra of many left- and right-going scans for 33 minutes of observation at 150 GHz. For this period, as reported by MERRA-2, there was a northwesterly wind at $h=1$\,km with speed $v_w=25$\,m/s and bearing $\phi_w=293^\circ$ while ACT was scanning centered at azimuth $\phi = 40^\circ$. The thin lines show the spectra for each 60$^\circ$-wide scan, and the thick lines show the median spectrum for each direction. The inset shows the estimated array-relative atmospheric velocity, which explains the difference in the spectra. Left-going scans correspond to a northwest-going motion tend to move through the atmosphere more quickly, which causes their spectrum to shift to the right and the $1/f$ knee frequency to increase. The opposite is true for right-going scans.}
\label{fig:scan_shift}
\end{figure}

Knowledge of the wind speed can improve our understanding of the atmospheric contribution to the noise during CMB observations.
As atmospheric brightness fluctuations are driven by the inhomogenous distribution of water vapor moving through the line-of-sight of the telescope, an increase in the relative velocity at which those distributions move across the array will affect the resulting time-ordered spectrum. Consider a telescope pointing due north while wind moves the atmosphere from west to east. Left-going (counterclockwise) scans will have a net west-going velocity and will thus scan ``against" the atmosphere, while right-going scans will analogously scan ``with" the atmosphere. This leads to a scan asymmetry, where the left-going scans will measure the atmosphere as moving relatively faster than right-going scans, leading to differing properties of the time-wise spectrum of the data. Moving the atmosphere more quickly through a beam has the effect of shifting its power spectrum toward higher frequencies, and due to the approximately scale-invariant angular power spectrum of the atmosphere, this is roughly equivalent to scaling the entire spectrum by some constant. The phenomenon is illustrated in Figure~\ref{fig:scan_shift}. 

\begin{figure}[htp]
\includegraphics[width=1\columnwidth]{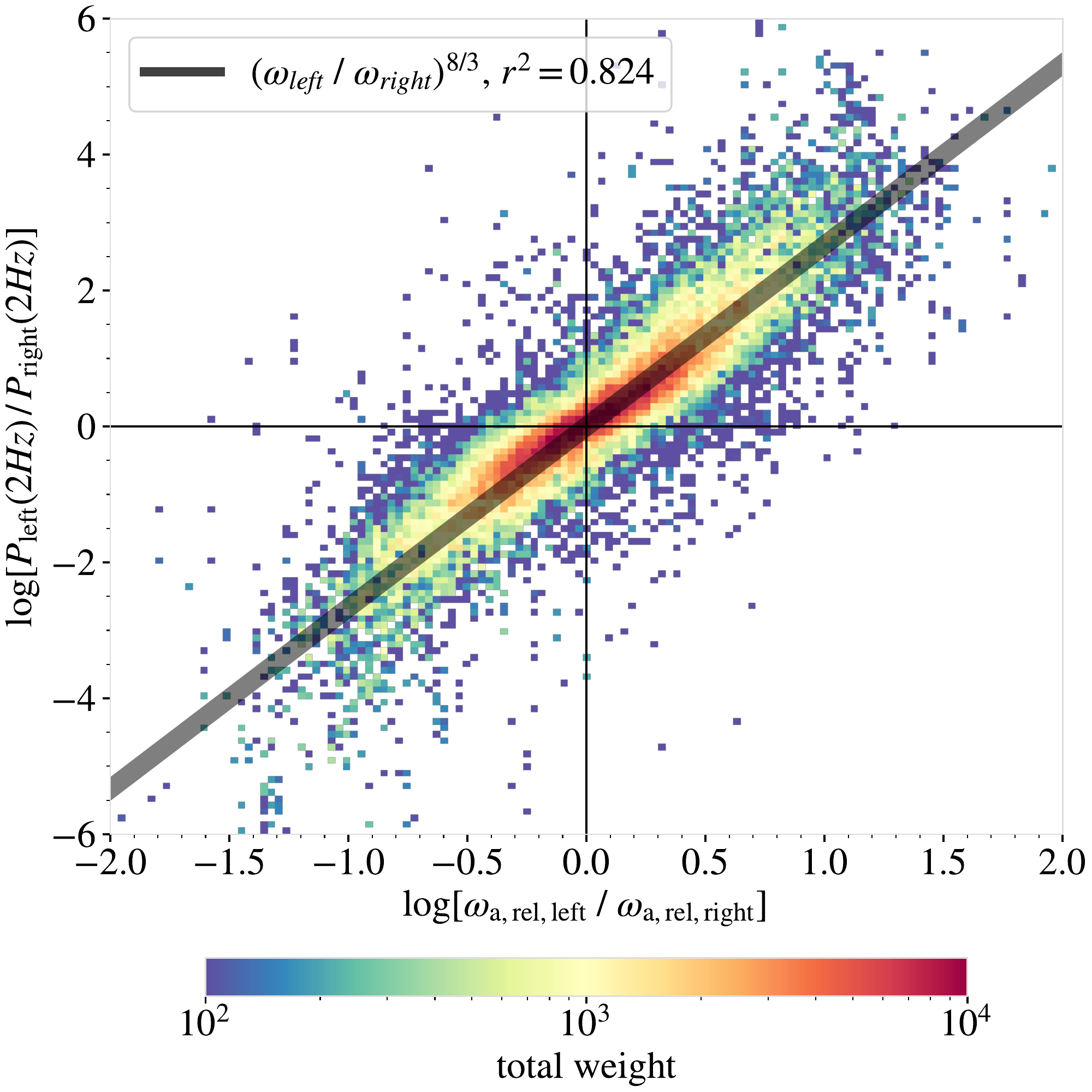}
\caption{A representation of Equation~\ref{eqn:theo_imb_power} for ACT data showing the determination of the log-ratio of the left-going and right-going power at 2\,Hz (estimated by averaging from 1--3 Hz), where the spectrum is almost always atmosphere-dominated. The total weight is given by the sum of the model's left-going weight and right-going weight. The log-power ratio is fairly well predicted by the log-speed ratio, with a weighted coefficient of determination of $r^2=0.824$ and with the expected log-space slope of $b=8/3$.}
\label{fig:imbalance_power}
\end{figure}

\begin{figure}[htp]
\includegraphics[width=1\columnwidth]{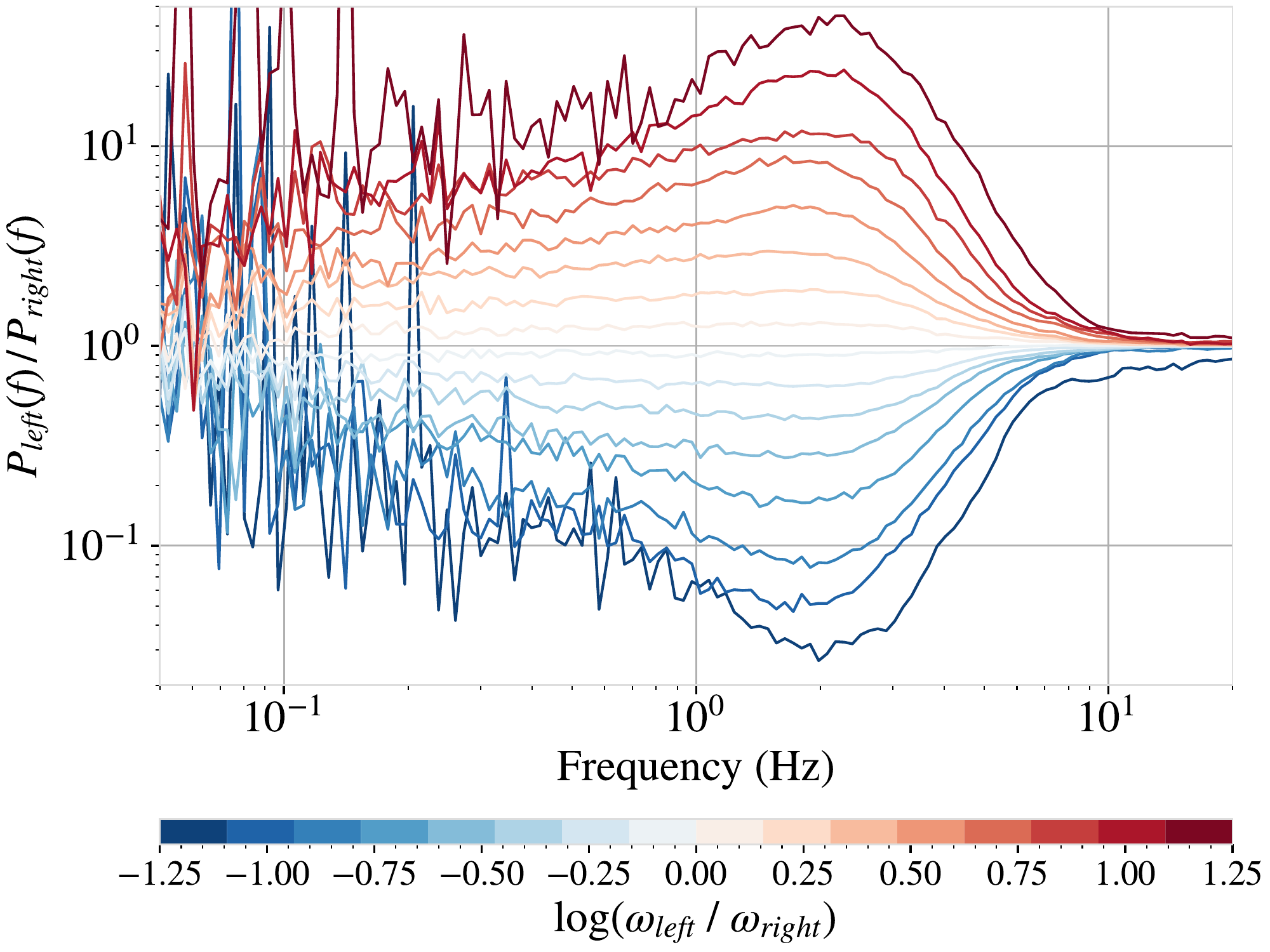}
\caption{A plot of the ratio of left-going and right-going atmospheric spectra (Figure~\ref{fig:scan_shift}).
The lines are color-coded to indicate the velocity bin and binned by log-ratio of left-going velocity and right-going velocity for around 7200 hours of data PA6 at 150 GHz. Depending on the orientation of the telescope, a right-going scan can be with or against the wind.}
\label{fig:imbalance_spectra}
\end{figure}

We find that, in general, the array-relative atmospheric motion (as computed for each stretch of data in the previous section) is a good predictor of the asymmetry in left-going and right-going power spectra of ACT data. In particular, for a scale-invariant spectrum we have

\begin{equation}
\label{eqn:theo_imb_power}
    \log \left [  \frac{|\vec{\omega}_\mathrm{a, rel, left}|}{|\vec{\omega}_\mathrm{a, rel, right}|} \right ] \propto -b \log \left [ \frac{P(f)_\mathrm{left}}{P(f)_\mathrm{right}} \right ],
\end{equation}
where $\vec{\omega}_\mathrm{a,rel, left}$ and $P(f)_\mathrm{left}$ are the aggregate relative atmospheric velocity and power spectrum for left-going scans (and analogously for right-going scans), and $b$ is the index of the power spectrum. Approximately 96\% of estimated scanning motion log-ratios are between $-1$ and $+1$, and the measured atmospheric power ratios follow the expectation from the model as shown in Figure~\ref{fig:imbalance_power}. Figure~\ref{fig:imbalance_spectra} shows the ratios of left-going and right-going spectra from full (not subdivided) scans, discriminated by the ratio of the magnitude of the left-going and right-going atmospheric velocities for 4000 hours of observation by PA6.\footnote{The left- and right-going velocities for each TOD are given by the weighted mean of all scans in that TOD, where the weights are determined as in the previous section.} It is not uncommon for the power spectra of the different directions to differ by almost an order of magnitude. 

Knowledge of the array-relative velocities can be used to build models that minimize the effects of atmospheric fluctuations in the data. These array-relative velocities can be directly computed from the data using the pair-lag method, and are related to the wind velocity as
%
\begin{equation}
\vec{\omega}_\mathrm{a} = 
\begin{cases}
\omega_{\mathrm{a},x} = -(u_\mathrm{east}\cos\phi - u_\mathrm{north}\sin\phi)\sin\epsilon + \frac{d\phi}{dt}\cos\epsilon\\
\omega_{\mathrm{a},y} = -(u_\mathrm{east}\sin\phi + u_\mathrm{north}\cos\phi)\sin^2\epsilon - \frac{d\epsilon}{dt} ,
\end{cases}
\end{equation}
where $\dot\phi = -1.5 $\,deg\,/\,s for left-going scans and $1.5 $\,deg\,/\,s for right-going scans. The asymmetric spectrum between scan directions is most pronounced when $\phi$ and $\phi_u = \tan^{-1}[u_\mathrm{east} / u_\mathrm{north}]$ are orthogonal, and minimized when they are parallel. 


\section{Discussion and Conclusion}

We have presented a method for deriving the angular wind velocity on ten-second time scales using data from ACT detector arrays without any other input. The method works by solving for the speed and direction of the frozen-in small-scale turbulent distribution of water vapor as it traverses the arrays. By averaging the derived wind velocity over an hour we can compare ACT to external weather sources like APEX, ERA5 and MERRA-2. To compare to APEX, we connect ACT and APEX measurements with an effective scale height. To compare to ERA5 and MERRA-2, we develop a model that uses their three-dimensional distribution of temperature and water vapor to predict the angular wind velocity as seen by ACT. The agreement between all four is quite good, suggesting that our physical picture of atmospheric emission resembles reality. Our investigation also shows good agreement of the PWV between ACT, APEX, ERA5, MERRA-2, and \citet{cortes/etal:2020}. Further adaptations of the pair-lag method to telescopes with different optical characteristics located in different geographical sites will help to better understand the motion-driven emission fluctuations of the atmosphere. This work is generalizable, with some adjustment, to any millimeter-wave telescope that observes the CMB with multiple detectors. Estimating the velocity of the atmosphere relative to ACT is also a good first-order predictor of the difference in the noise properties between left- and right-going scans, which can be quite substantial.


As our ability to understand and model atmospheric fluctuations improves, we hope to be able to probe the CMB temperature anisotropy to larger and larger angular scales (lower $\ell$). In addition to enhancing our ability to calibrate to {\sl Planck} \citep[e.g.,][]{hajian/etal:2011}, it will improve ACT's ability to investigate cosmology independent of {\sl Planck} and {\sl WMAP}. In particular, pushing to larger scales should improve the TE correlation, an especially effective spectrum for constraining cosmology, and the TB correlation which is an important check of systematic errors. 

\section{Acknowledgments} 

This work was supported by the U.S. National Science Foundation through awards AST-0408698, AST-0965625, and AST-1440226 for the ACT project, as well as awards PHY-0355328, PHY-0855887 and PHY-1214379. 
Funding was also provided by Princeton University, the University of Pennsylvania, and a Canada Foundation for Innovation (CFI) award to UBC. 
ACT operates in the Parque Astron\'omico Atacama in northern Chile under the auspices of the Agencia Nacional de Investigaci\'on y Desarrollo (ANID).
TM gratefully acknowledges financial support through Prof. David Spergel. The work was also supported by the Misrahi and Wilkinson funds and made use of the Della computer cluster. RB acknowledges support for the UdeC-UCSC 183 GHz radiometer from the UCSC project DINREG 06/2017. SKC acknowledges support from NSF award AST-2001866. ADH acknowledges support from the Sutton Family Chair in Science, Christianity and Cultures and from the Faculty of Arts and Science, University of Toronto. ZX is supported by the Gordon and Betty Moore Foundation.

We gratefully acknowledge the many publicly available software packages that were essential for parts of this analysis including \texttt{numpy}, \texttt{scipy}, and \texttt{scikit-learn}.
We also acknowledge use of the \texttt{matplotlib}~\citep{Hunter:2007} package and the Python Image Library for producing plots in this paper.

\bibliographystyle{wmap}
\bibliography{_refs.bib,apj-jour}

\onecolumngrid
\appendix
\section{Structure function for overlapping beams}
\label{appen:a}

\begin{figure*}
\epsscale{1} 
\plotone{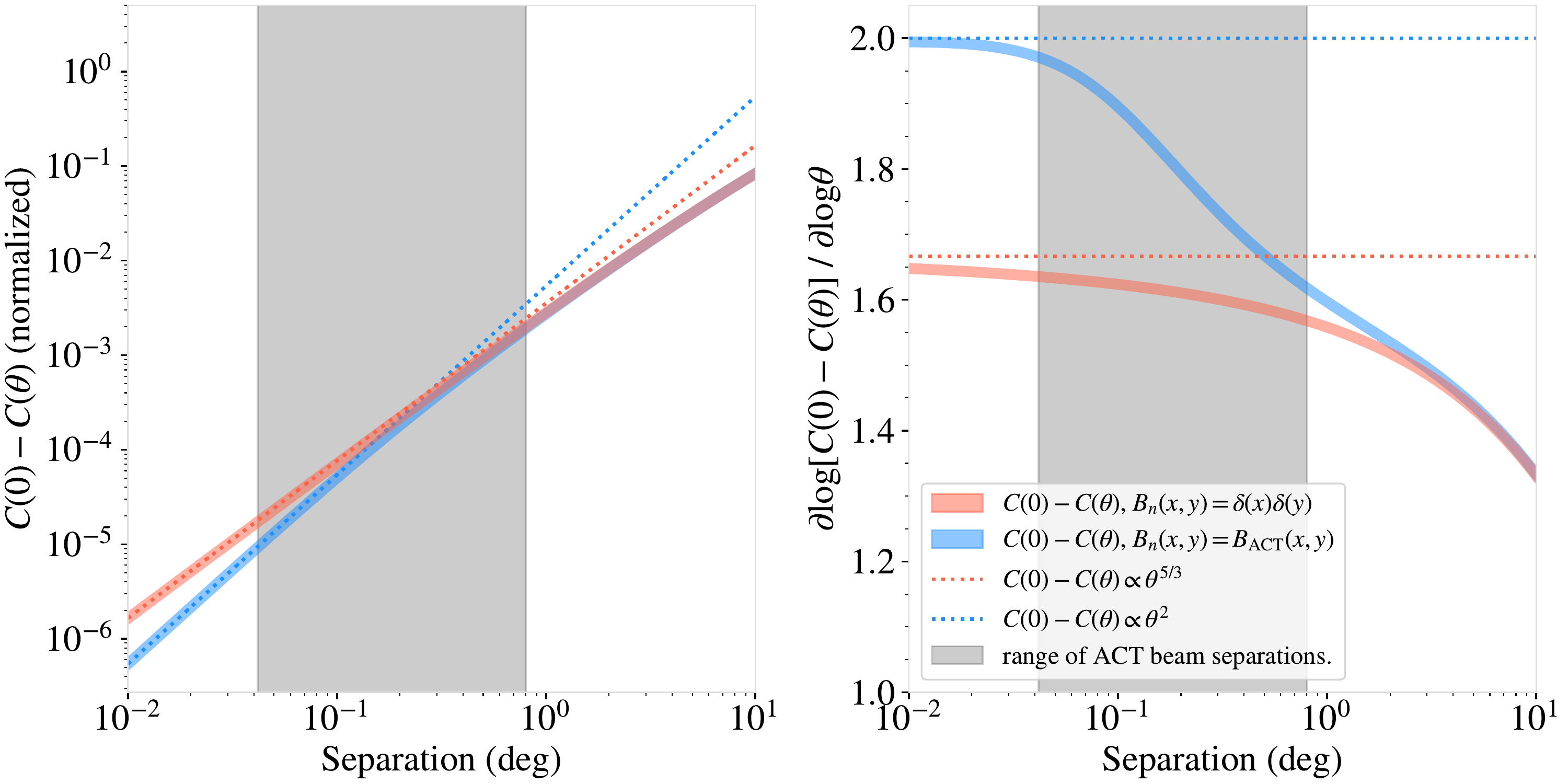}
\caption{Monte-Carlo integration (iterated until errors became negligible) of Equation~\ref{eqn:appendix_beam_corr} for separations between $0.01^{\circ}$ degrees and $10^{\circ}$, for thin beams (red) and ACT-like beams (blue). Power law structure functions with indices of 5/3 (red dotted) and 2 (blue dotted) are superimposed. Angular separations between detectors on the ACT focal plane are between $2.4'$ and $0.8^\circ$; this paper focuses on that regime. }
\label{fig:monte_carlo_beams}
\end{figure*}

This appendix presents the derivation of the structure function in Equation~\ref{eqn:powlawtwo}. 
We consider a three-dimensional distribution of atmospheric water vapor that follows the three-dimensional turbulent correlation $D(r)$, scaled exponentially as a function of height above the ground. We start with Equation~\ref{eqn:six_integral_3}: 
\begin{equation}
\label{eqn:appendix_beam_corr}
    C(\theta )= \alpha_b(\nu)^2 \rho(0)^2 T_\mathrm{atm}(0)^2 \iiint \iiint D(r_{ij}) e^{-(z_i + z_j)/z_0} B_{n,i}(x_i,y_i) B_{n,j}(x_j,y_j) dx_idy_idz_i dx_jdy_jdz_j,
\end{equation}
with atmospheric correlation $D(r)$ and a beam 
function $B_n(x,y)$ that is roughly constant in $z$
as described in Section~\ref{sec:angcorr}.
The explicit physical distance $r_{ij}$ between two atmospheric elements $dx_idy_idz_i$ and $dx_jdy_jdz_j$ for two beams separated by angle $\theta$ is given by the expression (see Figure~\ref{fig:sep_frame})
\begin{equation}
\label{eqn:appendix_geom}
    r_{ij} = \Big [ \big ( (x_i - x_j)\cos(\theta/2) + (z_i+z_j)\sin(\theta/2) \big )^2 + \big ( y_i - y_j \big )^2 + \big ( (z_i - z_j)\cos(\theta/2) + (x_i+x_j)\sin(\theta/2) \big )^2 \Big ] ^{1/2},
\end{equation}
which for small $\theta$ may be written as
\begin{equation}
    r_{ij} = \Big [ \big ( x_i - x_j + (z_i+z_j)(\theta/2) \big )^2 + \big ( y_i - y_j \big )^2 + \big ( z_i - z_j \big )^2 \Big ] ^{1/2},
\end{equation}
where because the beams are much longer in $\hat{z}$ than they are wide, we drop the $(x_i+x_j)\sin(\theta/2)$ term but keep the $(z_i+z_j)\sin(\theta/2) \approx (z_i+z_j)(\theta/2)$ term. Changing to an integration over variables $z = (z_i + z_j) / 2$ and $z_\Delta = z_i - z_j$, the full expression becomes
\begin{equation}
\label{eqn:integrate_var_sub}
     C(\theta) = A \iiiint \int_0^\infty \int_0^{2z} D \Big [ \big ( ( x_i - x_j + z\theta  ) ^ 2 + ( y_i - y_j ) ^ 2 + z_\Delta^2 \big )^{1/2} \Big ]  e^{-2z/z_0} B_{n,i}(x_i,y_i) B_{n,j}(x_j,y_j) dz_\Delta dz dx_idy_i dx_jdy_j, 
\end{equation}
where $A=\alpha_b(\nu)^2 \rho(0)^2 T_\mathrm{atm}(0)^2$ and has units of $\mathrm{K}^2 \mathrm{m}^{-2}$. We focus on the integral over $z_\Delta$. Consider the covariance element 
\begin{equation}
    dC(\theta, \b{r}_5 ) = A \int_0^{2z} D \Big [ \big ( ( x_i - x_j + z\theta  ) ^ 2 + ( y_i - y_j ) ^ 2 + z_\Delta^2 \big )^{1/2} \Big ] dz_\Delta,
\end{equation}
where $\b{r}_5$ represents a specific 5-tuple of coordinates $(x_i,y_i,x_j,y_j,z)$. We write this quantity as 
\begin{equation}
dC(\theta, \b{r}_5 ) = A \int_0^\infty D \Big [ \big ( ( x_i - x_j + z\theta  ) ^ 2 + ( y_i - y_j ) ^ 2 + z_\Delta^2 \big )^{1/2} \Big ]  dz_\Delta - A \int_{2z}^\infty D \Big [ \big ( ( x_i - x_j + z\theta  ) ^ 2 + ( y_i - y_j ) ^ 2 + z_\Delta^2 \big )^{1/2} \Big ]  dz_\Delta. 
\label{eqn:integrate_var_sub2}
\end{equation}

The term on the right varies negligibly in $\theta$ when $\theta$ is small as we always have $z_\Delta \gg z\theta$, and thus we take it as a constant element $dB(\b{r})$. The term on the left is more illuminating. We introduce the quantity $\theta_{\mathrm{eff}}$, defined such that
\begin{equation}
    z\theta_{\mathrm{eff}} = \big ( (x_i - x_j + z\theta  ) ^ 2 + ( y_i - y_j ) ^ 2 \big )^{1/2},
\end{equation}
which is constant for the integral over $z_\Delta$. Plugging in the atmospheric correlation function $D(r)$ (Equation~\ref{eqn:matern}) into Equation~\ref{eqn:integrate_var_sub2} gives us 
\begin{equation}
\label{eqn:cov_element}
    dC(\theta, \b{r}_5 ) = \frac{2^{2/3}}{\Gamma \big ( \frac{1}{3} \big )} A r_0^{-1/3} \int_0^\infty  \Big (z^2\theta_{\mathrm{eff}}^2+z_\Delta^2\Big )^{1/6} K_{1/3} \Big [ r_0^{-1} \big ( z^2\theta_{\mathrm{eff}}^2+z_\Delta^2 \big )^{1/2} \Big ] dz_\Delta + dB(\b{r}_5).
\end{equation}

Now consider the identity~\citep{bateman}:
\begin{equation}
    \mathcal{F}^{-1} \Big [(x^2+b^2)^{\nu/2} K_{\nu}\big [a (x^2 + b^2)^{1/2} \big ] \Big ] (\gamma) = a^{\nu}b^{\nu+1/2}(\gamma^2+a^2)^{-\nu/2-1/4} K_{-\nu-1/2}\big [ b (\gamma^2 + a^2 )^{1/2} \big ],
\end{equation}
which holds when $a$ and $b$ are strictly positive. Setting $a = zr_0^{-1}$, $b=z^{-1}z_\Delta$, $\nu=1/3$, and equating $x$ with $\theta_\mathrm{eff}$ allows us to write the covariance element as
\begin{equation}
      dC(\theta, \b{r}_5 ) = 
      \frac{2^{2/3}}{\Gamma \big ( \frac{1}{3} \big )} A r_0^{-2/3} z^{-1/6} \mathcal{F} \bigg[ 
       (\tilde{\theta}_{\mathrm{eff}}^2+z^2r_0^{-2})^{-5/12} \int_0^\infty z_\Delta^{5/6}  K_{-5/6}\big [ z^{-1} z_\Delta (\tilde{\theta}_{\mathrm{eff}}^2 + z^2r_0^{-2} )^{1/2} \big ] dz_\Delta \bigg] + dB(\b{r}_5) 
\end{equation}
\begin{equation}
\label{eqn:integral_spectrum}
      = \frac{(2\pi)^{1/2} \Gamma \big ( \frac{4}{3} \big )}{\Gamma \big ( \frac{1}{3} \big )} A z^{5/3} r_0^{-2/3}
       \mathcal{F} \bigg[ (\tilde{\theta}_{\mathrm{eff}}^2+z^2r_0^{-2})^{-4/3} \bigg] + dB(\b{r}_5), 
\end{equation}
where $\tilde{\theta}_{\mathrm{eff}}$ is the Fourier conjugate of $\theta_{\mathrm{eff}}$. We can 
evaluate Equation~\ref{eqn:integral_spectrum} to yield 
\begin{equation}
     dC(\theta, \b{r}_5 )= \frac{(2\pi)^{1/2} \Gamma \big ( \frac{4}{3} \big ) }{\Gamma \big ( \frac{1}{3} \big )} A z^{5/3} r_0^{-2/3}
       \bigg[ \frac{1}{2^{1/3} \Gamma \big ( \frac{4}{3} \big ) } \big ( \theta_{\mathrm{eff}} r_0 z^{-1} )^{5/6} K_{5/6}\big[z\theta_{\mathrm{eff}} r_0^{-1} \big ]   \bigg] + dB(\b{r}_5)
\end{equation}
\begin{equation}
\label{eqn:single_integral}
      =  \frac{2^{1/6}\pi^{1/2}}{\Gamma \big ( \frac{1}{3} \big ) } A r_0
         \big ( z\theta_{\mathrm{eff}} r_0^{-1} \big )^{5/6} K_{5/6}\big[z\theta_{\mathrm{eff}} r_0^{-1} \big ] + dB(\b{r}_5).
\end{equation}

We now have
\begin{multline}
\label{eqn:five-integral}
    C(\theta) = \iiiint \int_0^\infty dC(\theta, \b{r}_5 ) B_{n,i}(x_i,y_i) B_{n,j}(x_j,y_j) dz dx_idy_i dx_jdy_j \\ = \frac{2^{1/6}\pi^{1/2}}{\Gamma \big ( \frac{1}{3} \big ) } A r_0 \iiiint \int_0^\infty (z\theta_{\mathrm{eff}} r_0^{-1})^{5/6} K_{5/6}\big[z\theta_{\mathrm{eff}} r_0^{-1} \big ] e^{-2z/z_0} B_{n,i}(x_i,y_i) B_{n,j}(x_j,y_j) dz dx_idy_i dx_jdy_j + B,
\end{multline}
where 
\begin{equation}
    B = \iiiint \int_0^\infty dB(\b{r}_5) B_{n,i}(x_i,y_i) B_{n,j}(x_j,y_j) dz dx_idy_i dx_jdy_j.
\end{equation}

Consider the special case $B_n(x,y) \to \delta(x)\delta(y)$, which describes the beam function in the limit of an infinitely thin cylinder. In this case,  $\theta_\mathrm{eff} \to \theta$ and the above expression reduces to
\begin{multline}
\label{eqn:single_integral2}
    C(\theta) = \iiiint \int_0^\infty dC(\theta,\b{r}_5) dz \delta(x_i)\delta(y_i) \delta(x_j)\delta(y_j) dx_idy_i dx_jdy_j \\ = \frac{2^{1/6}\pi^{1/2}}{\Gamma \big ( \frac{1}{3} \big ) } A r_0  \int_0^\infty (z\theta r_0^{-1})^{5/6} K_{5/6}\big[z\theta r_0^{-1} \big ] e^{-2z/z_0} dz + B.
\end{multline}
When $\theta \ll z^{-1}r_0 $ for all $z$, we may approximate the proportionality of the structure function $C(0) - C(\theta)$ as
\begin{equation}
\label{eqn:pl_sf}
    C(0) - C(\theta) \propto \int_0^\infty (z\theta r_0^{-1})^{5/3} e^{-2z/z_0} dz \propto \theta^{5/3},
\end{equation}
where $B$ drops out due to its negligible dependence on $\theta$. Note that the relative contribution of the layers to the structure function decreases twice as fast as the water vapor scaling. We can approximate this integral with a sum over discrete layers of angle-dependent emission at variable distance $z$ along the beam, where each has a $5/3$ structure function in $\theta$. In reality, beams do not have infinitely small waists. Realistically treating the beam geometry requires us to explicitly compute the five-integral in Equation~\ref{eqn:five-integral}, which is difficult to do analytically. A more expedient approach is to compute it numerically; fortunately, computing the normalized structure function does not require us to compute either $A$ or $B$.

Figure~\ref{fig:monte_carlo_beams} shows the result of stochastically computing the normalized angular atmospheric structure function for very thin beams (negligible width) and ACT-like beams (5.5 meters wide) using a Monte-Carlo method, iterated until errors became negligible. We see that for the thin beams approximation, we recover the expected $5/3$-index for the structure function for small separations. However, we see that the structure function of the atmosphere as seen by ACT is better approximated by an index of between $1.6$ and $2$ for small separations. In both cases, the slope of the structure function decreases for larger separations as the outer scale of turbulence becomes non-negligible. We use this to justify the least-squares solution for the atmospheric motion as seen by ACT, and also to justify the layered two-dimensional structure function of the atmosphere in Equation~\ref{eqn:single_sum}; despite the deviation at larger separations of the full integral from the index of 2 used in Eq. 21, the results in this paper show that the constant-index approximation works remarkably well at modelling the motion of the atmosphere. We also note that for separations larger than a degree, beam geometries become negligible.

\end{document}